\documentclass[aps,prd,preprintnumbers,nofootinbib,floatfix,11pt]{revtex4-2}
\pdfoutput=1

\usepackage{amsmath}
\usepackage{amssymb}
\usepackage{braket}
\usepackage{xcolor}
\usepackage{graphicx}
\usepackage{subfigure}
\usepackage{comment}
\usepackage{hyperref}
\usepackage{nameref}

\usepackage{amsthm}

\newcommand{\mykappa}{{\kappa}}

\newcommand{\myvar}{{\varphi}}

\begin{document}

\title{Squeezed States in Gravity}
\author{Arunima Das$^{1}$, Maulik Parikh$^{1,2}$, Frank Wilczek$^{1,3,4,5}$, and Raphaela Wutte$^{1,2,6}$}

\affiliation{$^{1}$Department of Physics, Arizona State University, Tempe, Arizona 85287, USA}
\affiliation{$^{2}$Beyond Center for Fundamental Concepts in Science, Arizona State University, Tempe, Arizona 85287, USA} 
\affiliation{$^{3}$Department of Physics, Stockholm University, Stockholm SE-106 91, Sweden}
\affiliation{$^{4}$Center for Theoretical Physics, Massachusetts Institute of Technology, Cambridge, Massachusetts 02139, USA}
\affiliation{$^{5}$Wilczek Quantum Center, Department of Physics and Astronomy, Shanghai Jiao Tong University, Shanghai 200240, China}
\affiliation{$^{6}$STAG Research Centre and Mathematical Sciences, University of Southampton, Highfield, SO17 1BJ Southampton, UK}
\begin{abstract}
\begin{center}
{\bf Abstract}
\end{center}
\noindent
We present a general framework for the production of squeezed quantum states of the gravitational field in linearized quantum gravity. Time-dependent couplings in the quadratic part of the action generically produce squeezed states from the vacuum. Using the harmonic oscillator as an example, we describe three techniques to obtain the squeezing parameter from such quadratic terms. For gravity, the action to quadratic order in metric perturbations contains couplings both to background curvature as well as to matter sources. Thus, both time-dependent classical spacetimes and time-dependent classical matter typically produce squeezed states of gravity.
\end{abstract}
 
\maketitle  
    

\newpage
\section{Introduction}

Coherent states are, in a sense, the most classical of quantum states, enjoying the minimum uncertainty required by the uncertainty principle. For the harmonic oscillator, the position-space wavefunction is a Gaussian whose midpoint traces the motion of the classical oscillator, while preserving its shape at all times. But while coherent states indeed have classical features, they are not the only ones: squeezed coherent states (sometimes called displaced squeezed states) also have minimum uncertainty and wavefunctions that do not disperse over time. Thus there is more than one possibility for the quantum state corresponding to a system that behaves classically. In particular, while a hundred years of observation have shown the gravitational field to be in perfect agreement with the classical equations of Einstein's theory, its quantum state remains undetermined.

Now, in order to show that the state not only {\em could} be squeezed, but {\em had} to be squeezed, one needs to have a theory of how states are ``produced" i.e. how interactions with sources (or self-interactions) cause a state that is initially a vacuum state to evolve into something nontrivial. This depends on the nature of the couplings in the theory. For electrodynamics, linear couplings in the Maxwell action guarantee that classical sources governed by deterministic dynamics produce coherent states of the electromagnetic field. This is easy to see in interaction picture, where the linear coupling $\int J A$ leads, after $A$ is expanded in creation and annihilation operators, to a state 
$|\Psi \rangle \sim \exp [-i \int dt dk  \tilde{J}(k,t)  a_k^\dagger ]|0 \rangle$ i.e. a coherent state (for a synopsis of coherent and squeezed states, see the appendix). Thus classical sources produce field states that most resemble the classical solutions of Maxwell's equations. Deviating from classicality in electrodynamics is not straightforward: for example, squeezed states of the electromagnetic field can be produced via parametric down-conversion in an optical cavity, thereby creating an effectively nonlinear interaction.

Gravity is different. A perturbative expansion of the gravitational action --- with or without matter sources and even before taking into account higher-curvature corrections to the Einstein-Hilbert action --- contains terms of all positive powers of the metric perturbation. By expanding these metric perturbations in creation and annihilation operators, we obtain states that are exponentials of polynomials of such operators acting on the vacuum. Such states, which we have termed molded states, are not very tractable. However, if we focus on the leading nonlinear terms in the perturbation, we obtain a corresponding interaction Hamiltonian that is quadratic in creation and annihilation operators. But crucially (and unlike electrodynamics), the quadratic terms inherently contain couplings to the background, both because $\sqrt{g} g^{ab} R_{ab}$ contains three terms that can be expanded perturbatively and because commutators of derivatives introduce curvature.

Here we will analyze the gravitational states produced by time-dependence in the quadratic terms of the matter-gravity action: we will see that they correspond to squeezed states \cite{Glauber:1963fi,Glauber:1963tx} of the gravitational field. The quadratic terms are of two types. First, there could be time-dependence in the spacetime curvature; examples include cosmological backgrounds and black hole mergers. Second, even in a flat or stationary background, time-dependence of the sources (whose backreaction has been neglected) typically also produces squeezed states when the matter action is expanded to second order in metric perturbations. In gravity, there is always nonlinear coupling to matter, because both the inverse metric $g^{ab}$ and $\sqrt{-g}$ have non-terminating expansions in powers of the metric perturbation. That noncoherent states of gravity are so easily produced is quite striking.

Squeezed states of gravity are of more than formal interest. They had been considered long ago in the context of inflation \cite{Grishchuk:1975uf,Grishchuk:1989ss,Grishchuk:1990bj,Albrecht:1992kf}, and more recently in the context of black hole inspirals and ring-downs \cite{Guerreiro:2025sge,Manikandan:2025dea,Guerreiro:2025mcu,Kanno:2025how}. They are a particularly interesting class of states because they lead to fundamental noise at gravitational wave detectors \cite{Parikh:2020nrd,Parikh:2020fhy,Parikh:2020kfh,Kanno:2020usf,Cho:2021gvg,Haba:2020jqs,Parikh:2023zat,Hertzberg:2021rbl,McCuller:2021mbn,Jarzabczyk:2026} that is exponentially enhanced in the squeezing parameter, relative to the noise in coherent states (though see \cite{Carney:2024dsj,Carney:2024wnp}). More generally, departures from the coherent state for gravity have experimental implications \cite{Manikandan:2025hlz,Manikandan:2025lfx,Manikandan:2025qgv} for a range of experimental configurations with counterparts in quantum optics.

Here we will present the general framework underlying the production of squeezed states in quantum gravity. While several of the specific cases and many of the equations have appeared before, there has not yet been a systematic study of how such states arise. We will work in the regime of perturbative quantum gravity, in which the question of what the quantum states of gravity are is free of considerations of background independence. In this regime, the quantum states of gravity are just those of a helicity-two field living in curved spacetime. Moreover, the perturbations can be expanded in terms of modes, each of which behaves, at leading order, as a harmonic oscillator. Thus we begin, appropriately, with a review of squeezed states of the harmonic oscillator. We will focus especially on extracting the squeezing parameters from time-dependent quadratic terms in the action. We do this in three ways: by expressing the time-evolution equations of the squeezing parameters in a form suitable for numerical treatment, by relating the squeezing parameters analytically to Bogolyubov coefficents derived from solutions to the equations of motion, and by using group theory to extract the squeezing parameters from the Magnus expansion in perturbation theory. We then briefly comment on extensions to field theory. We then turn to gravity and identify the terms in the gravitational action that can generate squeezed states of metric perturbations. As examples, we explicitly calculate the time-dependence of the squeezing parameter of perturbative gravitational modes for FRW spacetimes, for the Kasner universe, and for a particle on a Newtonian circular orbit in a Schwarzschild geometry.

\section{Harmonic Oscillator}
\label{sec:harm}
Consider a harmonic oscillator with unit mass perturbed by  a quadratic time-dependent potential:
\begin{equation}
\label{eq:Lagrangianharm}
    L = \frac{1}{2} {\dot x}^2 - \frac{1}{2} \omega_0^2 x^2 + F(t) x^2\,.
\end{equation}
If the perturbation had been of the linear form $F(t)x$, the equation of motion would have an inhomogeneous driving force $F(t)$. But because the coupling is quadratic, this is instead a parametrically-driven oscillator.
Defining the ladder operators of the unperturbed system
\begin{equation}
\label{quantize}
x= \sqrt{\frac{1}{2 \omega_0}} ( a+ a^\dagger)\,, \quad 
p = i \sqrt{\frac{\omega_0}{2 }}
( a^\dagger - a)\,,
\end{equation}
we have
\begin{align}
\label{hamiltonian}
   H 
    &=  \left(\omega_0  - \frac{F(t)}{\omega_0} \right)
     \left (a^\dagger a + \frac{1}{2} \right )  - \frac{F(t) }{2\omega_0} (a^2  + (a^\dagger)^2 )
\,.
\end{align}
Schumaker showed \cite{Schumaker:1986tlu} that the evolution operator
may be decomposed 
as 
\begin{equation}
\label{unitary}
U(t) = e^{i \delta - \frac{i}{2} \omega_0 t + \frac{i}{2 \omega_0} \int^t_0 F(t') dt'} 
\,D(\mu)\, 
S(r, \myvar) \,R(\theta) \,,
\end{equation}
where $D$, 
$S$ and $R$ are the (single-mode) 
displacement, 
squeezing and rotation operators: 
\begin{align}
    D(\mu) = \exp( \mu a^\dagger - \mu^* a)\,,
    \quad
    S(r, \myvar) = \exp \left( 
    \frac{1}{2} r( e^{- i \myvar} a^2 - e^{i \myvar} (a^\dagger)^2
    \right)\,, \quad
    R(\theta) = \exp(-i \theta a^\dagger a)
    \,.
\end{align}
Here $\myvar, \theta, r, \delta$ are real functions of time and $\mu$ is a complex function of time.
Using the identity
\begin{equation}
e^A B e^{-A} = B + [A,B] + \frac{1}{2}[A[A,B]] + ...
\end{equation}
we easily see that
\begin{subequations}
\label{SaS}
\begin{align}
    S^{\dagger} {a} S &= \cosh{r} \, {a} -  e^{i \myvar} \sinh{r} \, {a}^{\dagger}\,, & 
   S^{\dagger} {a}^{\dagger} S &= \cosh{r}\, {a}^{\dagger}  -  e^{-i \myvar} \sinh{r} \, {a}\,, \label{squeezeact} \\
      R^{\dagger} {a} R &= e^{-i \theta} {a}\,,  &
    R^{\dagger} {a}^{\dagger} R &= e^{i \theta} {a}^{\dagger} 
    \,, \label{rotact} \\
    D^{\dagger} a D &= a + \mu \,, & D^{\dagger} a^{\dagger} D &= a^{\dagger} + \mu^*\,.
\end{align}
\end{subequations}
Suppose that the system is initially in its unperturbed ground state until $t=0$ when the perturbation $F(t)$ is switched on. Then, in Schr\"odinger picture, the system evolves to 
$U(t)|0\rangle$. Since $a|0 \rangle = 0$, the rotation operator acts trivially on the vacuum; thus, upto an overall phase, the system is in a squeezed state.
We would like to determine the squeezing parameter $r(t)$ in terms of the function $F(t)$. We now describe three different approaches to calculating the squeezing parameters: by numerically evolving the time evolution equation, by using Bogolyubov transformations if the classical solutions are known, and by using a perturbative expansion if $F(t)$ has a small coefficient.

\subsection{Calculating the squeezing parameter I: time evolution}
\label{sec:timeevolution}
In the first approach, we insert \eqref{unitary} into $i \frac{d}{dt} U = H U$. Then, as shown in \cite{Schumaker:1986tlu}, the parameters must satisfy the following differential equations: 
\begin{subequations}
\label{schum}
\begin{align}
        &- \frac{\Dot{\myvar}}{2} + (\frac{\Dot{\myvar}}{2} + \Dot{\theta}) \cosh{2r} = \omega_0  - \frac{F(t)}{\omega_0}\,, \label{schu1}\\
      & -i \Dot{r} e^{i \myvar} + (\frac{\Dot{\myvar}}{2} + \Dot{\theta}) e^{i \myvar} \sinh{2r} =  - \frac{F(t)}{\omega_0} \,,\label{schumr}
        \\
         &\dot \delta
         + \frac{1}{2} \dot{\theta} - \omega_0  + \frac{F(t)}{\omega_0} = 0\,, \\ 
        &\dot \mu = -\mu \left(\dot r \tanh r+i \dot \theta \right)-e^{ i
   \myvar} \left(\dot{\mu}^* \tanh r + \mu^*  \left(\dot{r}+i (\dot \theta +\dot \myvar
   ) \tanh r \right)\right)\,.
\end{align} 
\end{subequations}
The last two equations govern the time evolution of the phase, $\delta(t)$, and the displacement parameter, $\mu(t)$. The latter is only present if there is a linear perturbation. 
Since we are interested in the squeezing from quadratic terms, we focus on the first two equations. These equations, which are coupled and nonlinear, cannot be solved analytically. They are also not well-suited for numerical 
evolution, because the phase $\myvar$ is not defined when $r = 0$, which is the initial condition we have at $t = 0$. Thus we need to re-write these in Cartesian coordinates on the squeezing parameter plane. Writing $\sinh(2r)e^{- i \myvar} = x + i y$, we find 
\begin{subequations}
\label{eqsfull1}
\begin{align}
   \dot x + 2 \omega_0 \left(1  -\frac{F}{\omega_0^2} \right) y &= 0\,\\
     \dot y - 2 \omega_0 \left(1-\frac{F}{\omega_0^2} \right) x &= \frac{2 F}{\omega_0} \sqrt{1+x^2+y^2}\,. 
     \label{schumakernumerics}
\end{align}
\end{subequations}
The equations can now be solved numerically, subject to the initial conditions
\begin{equation}
    x(0) = y(0) = 0\,.
\end{equation}
To get a sense of them, let us consider a few limiting cases.
From our classical intuition about parametrically driven oscillators, we expect an instability at $2 \omega_0$, the frequency of parametric resonance. Consider then an $F(t)$ of the form
\begin{equation}
\label{eq:Ftok}
    F(t) 
    = - \frac{A \epsilon}{4}\sin(2 \omega_0 t)\,.
\end{equation}
where $\epsilon$ is a small dimensionless parameter.
Let us solve Schumaker's equations perturbatively.
Expanding $x(t)$ and $y(t)$ as a power series in $\epsilon$,
and choosing the initial conditions 
   $x(0) = y(0) = 0$,
we find to first order in $\epsilon$ that
\begin{equation}
    x(t) = \frac{A \sin (2 \omega_0 t )}{8 \omega_0^2}-\frac{A \,
   t \cos (2 \omega_0 t )}{4 \omega_0}\,, \qquad 
   y(t) = -\frac{A \,t \sin (2 \omega_0 t )}{4 \omega_0}
\,,
\end{equation}
yielding 
\begin{equation}
r(t) = \frac{1}{2} \mathrm{arcsinh} \left(\frac{A}{4 \omega_0^2}
\left(\omega_0^2 t^2 - \omega_0 t \sin(2 \omega_0 t)
\cos(2 \omega_0 t) + \frac{1}{4} \sin^2(2 \omega_0 t)
\right)^{1/2}
\right)\,.
\end{equation}
We now also solve \eqref{schum}, \eqref{eq:Ftok}
for large $r(t)$ where $\cosh(2r) \sim \sinh(2r)$.
It is useful to split 
\eqref{schumr} 
into its real and imaginary parts.
Then, for large $r$, we find
\begin{equation}
\label{phiasymp}
     \omega_0 \frac{\dot \myvar}{2} + F(t) \cos(\myvar) = F(t) + \omega_0^2
\end{equation}
Expanding as, 
\begin{equation}
    \myvar(t) = \sum_{n=0}^\infty \epsilon^n \myvar_n(t) \,, \qquad 
    r(t) = \sum_{n=0}^\infty \epsilon^n r_n(t)\,,
\end{equation}
we readily obtain 
$\myvar_0(t)/2 = \omega_0 t +b $ with $b \in \mathbb{R}$
by solving \eqref{phiasymp} at zeroth order. Similarly, 
we obtain
$\dot r_0 = 0$ and 
\begin{equation}
    \dot r_1 = \frac{2 A}{8 \omega_0} \sin(2 \omega_0 t) \sin( \myvar_0)\,,
\end{equation}
which yields 
\begin{equation}
    r(t) = \tilde{r}_0 + \epsilon \left( c + \frac{A t}{8 \omega_0} \cos (2 b) - \frac{A}{32 \omega_0^2} \sin(2 (b+2\omega_0 t)) \right) + O(\epsilon^2)\,.
\end{equation}
Choosing the integration constants to be $\tilde{r}_0=0$, $b = 0$, $c= 0$, we have 
\begin{equation}
     r(t) =  \epsilon  \left(
    \frac{A t}{8 \omega_0}
    -\frac{A \sin (4 t 
    \omega_0)}{32 
    {\omega_0}^2}\right)\,.
\end{equation}
Thus, we see that the squeezing parameter grows linearly in time superimposed with a small oscillatory term.

\begin{figure}[h!]
    \centering
    \subfigure{
        \includegraphics[width=0.489\textwidth]{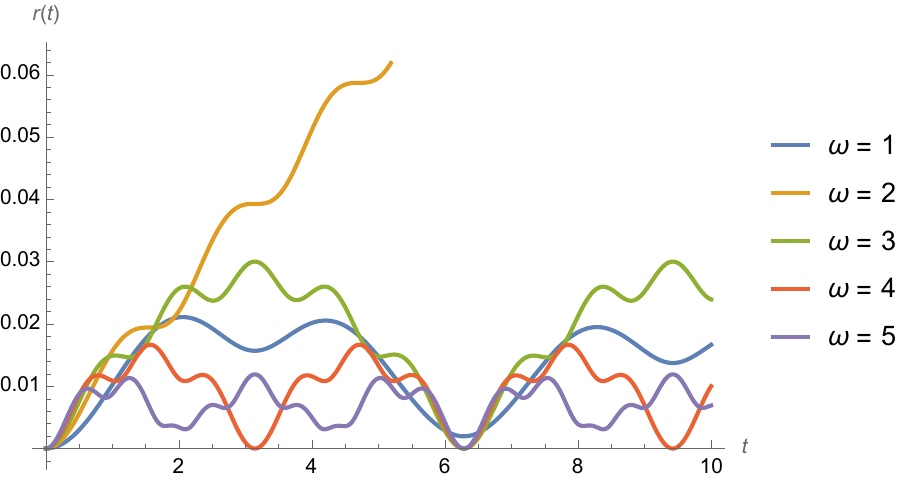}}
    \hfill
    \subfigure{
\includegraphics[width=0.489\textwidth]{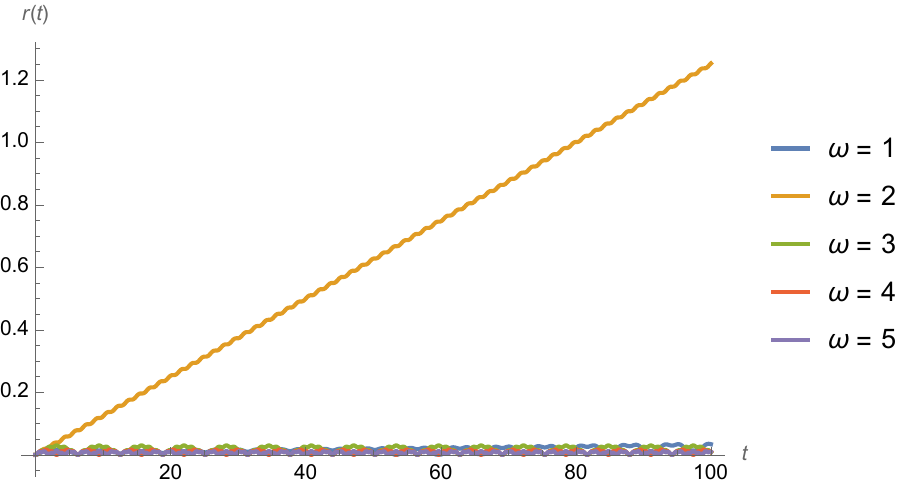}}
    \caption{Plot of the function $r(t)$ which can be inferred by numerically solving the equations \eqref{eqsfull1} with 
    $F(t) = -\frac{A}{4}\sin (2 \omega t)$ 
    with $A = 0.1$ and $\omega_0 =1$. 
    }
    \label{fig3}
\end{figure}

\begin{figure}[h!]
    \centering
    \subfigure{
        \includegraphics[width=0.43\textwidth]{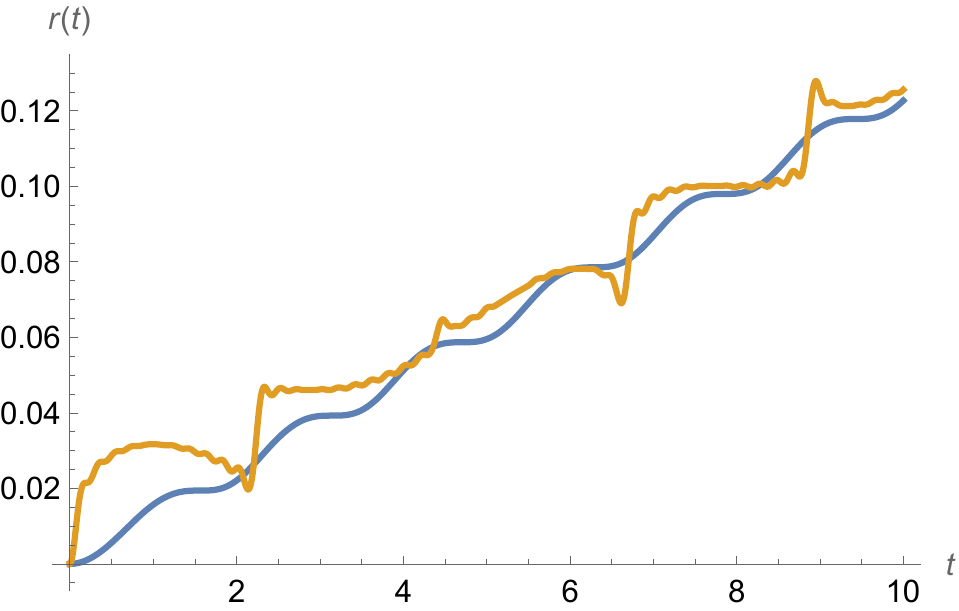}}
    \hfill
    \subfigure{
\includegraphics[width=0.43\textwidth]{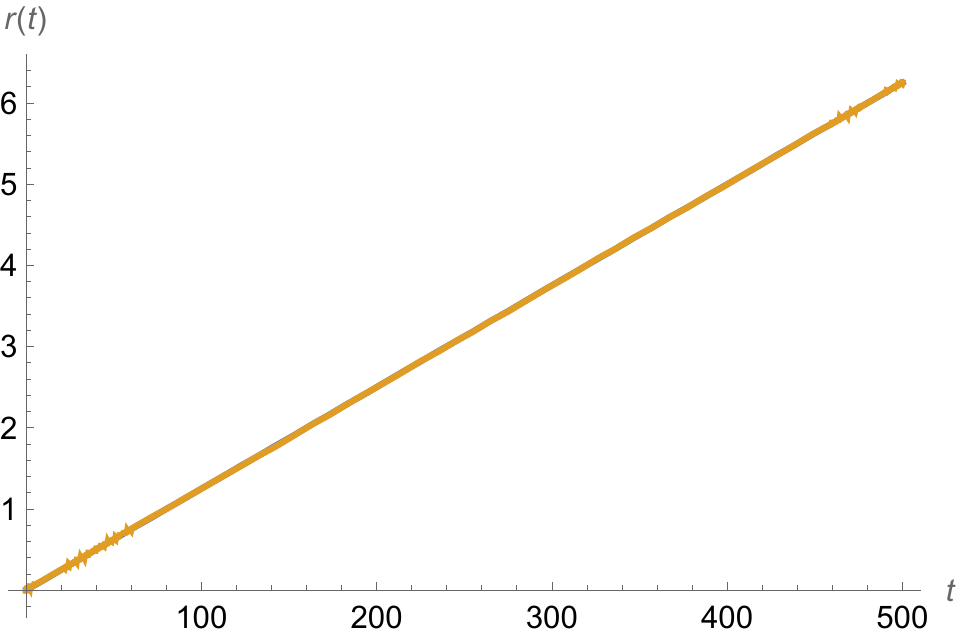}}
    \caption{Plot of the function $r(t)$ obtained by numerically solving the equations \eqref{eqsfull1} with 
    with 
    $F(t) = - \frac{A}{4} \sin(
    2 \omega_0 
    t)$ (in blue) and
    $F(t) = - \sum_{n = 0}^{10} \frac{A}{4}
    \sin(2 \omega_0 (1+ \sqrt{2} n) t)$ (in orange) with $A = 0.1$ and $\omega_0 =1$.
    }
    \label{fig4}

\end{figure}

Let us now consider some numerical solutions of \eqref{eqsfull1}. First consider $F(t) = -\frac{A}{4} \sin (2 \omega t)$, where $\omega$ is an integer multiple of $\omega_0$. Here we are driving the oscillator by higher overtones of $2 \omega_0$. The squeezing parameter is plotted in Fig. \ref{fig3} and indicates that only the driving term with frequency $2 \omega_0$ leads to secular growth in $r(t)$. 
The numerical solution approaches the asymptotic one for large times.
Finally, consider the effect of a function $F(t)$ of the form $F(t) = -\sum_{n = 0}^{N} \frac{A}{4}
    \sin(2 \omega_0 (1+ \sqrt{2} n) t)$, chosen as a superposition of functions whose frequencies are not rational multiples of each other. 
A plot for $N=10$ is given in Fig.\ \ref{fig4}. Because the equations are nonlinear, it is not the case that the contribution to the squeezing parameter of the different driving frequencies is additive. Nevertheless, the plot shows that the late-time behavior of the squeezing parameter is entirely dominated by the driving term at twice the fundamental frequency.

\subsection{Calculating the squeezing parameter II: Bogolyubov coefficients}
In the second approach, we look for solutions of the classical equations of motion. The squeezing parameters can then be obtained from the Bogolyubov coefficients of the solutions expressed in terms of the unperturbed solutions. This method is useful if the equations of motion can be solved in analytic form; to obtain the solutions one merely has to solve a linear differential equation rather than the coupled nonlinear ones of \eqref{schum}. It is also useful if we are only interested in the late-time behavior of the squeezing parameter, particularly if the solutions simplify in that limit. 

As before, we assume that $F(t) = 0$ for $t \leq 0$ and that the state of the system at $t = 0$ is the ground state of the harmonic oscillator with frequency $\omega_0$. 
Let ${\tilde \omega}^2 = \omega_0^2 - 2 F(t)$.
Suppose $u(t)$ is the (known) complex solution to the equation of motion
\begin{equation}
\ddot{u} + {\tilde \omega}^2(t) u = 0\,,
\end{equation}
satisfying the initial conditions 
\begin{equation}
u(0) = \frac{1}{\sqrt{2 \omega_0}} \,, \qquad \dot{u}(0) = -i \sqrt{\frac{\omega_0}{2}}\,.
\end{equation}

Then in Heisenberg picture we have 
\begin{equation}
x_H(t) =  a_0 u(t) + a^{\dagger}_0 u^*(t) \, , \qquad
p_H(t) =  a_0 \dot{u}(t) + a^{\dagger} \dot{u}^*
\end{equation}
Heisenberg operators obey the classical equations of motion and, indeed, we see that $x_H$ does that, since $u(t)$ is a solution. The state is $|\Psi \rangle$ where $a_0 |\Psi \rangle = 0$. Now define the operator
\begin{equation}
a_H(t) \equiv \sqrt{\frac{\omega_0}{2}} x_H(t) + i \sqrt{\frac{1}{2 \omega_0}} p_H(t)
\end{equation}
One has
\begin{equation}
a_H (t) = \alpha^*(t) a_0 - \beta^*(t) a_0^\dagger
\end{equation}
where
\begin{equation}
\label{alpha-beta}
\alpha(t) = \sqrt{\frac{\omega_0}{2}} u^*(t) - i \sqrt{\frac{1}{2 \omega_0}} \dot{u}^*(t) \qquad , \qquad \beta(t) = -\sqrt{\frac{\omega_0}{2}} u(t) + i \sqrt{\frac{1}{2 \omega_0}} \dot{u}(t)
\end{equation}
Of course, any two sets of ladder operators are related by a Bogolyubov transformation, with their corresponding ground states being related by the action of a squeezing operator. To show that the squeezing parameters are the {\em same} as the ones we defined earlier in \eqref{unitary}, we note that $a_H(t)$ is just the Heisenberg picture operator corresponding to $a_0$. Then $a_H(t) = U^\dagger a_0 U$, where $U$ is given by \eqref{unitary}. With $U \sim SR$, we can use \eqref{SaS} to write
\begin{equation}
a_H = e^{-i \theta} \cosh r a - e^{i (\theta + \varphi)} \sinh r a^\dagger
\end{equation}
from which we see that
\begin{equation}
r(t) =  \mathrm{arcsinh} |\beta(t)|\,,
\label{rbog}
\end{equation}
and 
\begin{equation}
e^{i \theta(t)} = \frac{\alpha(t)}{|\alpha(t)|} \qquad , \qquad e^{i \varphi(t)} = \frac{\beta^*(t)}{|\beta(t)|} \frac{|\alpha(t)|} {\alpha(t)} \, .
\end{equation}

\subsection{Calculating the squeezing parameter III: Magnus expansion}

In the third approach, we assume that the time-dependent part of the Hamiltonian has a small coefficient, $\lambda < 1$. We can then obtain the squeezing parameters perturbatively using the Magnus expansion and some group theory. Consider a Hamiltonian of the form $H_0 + \lambda H_{\rm int}(t)$, where $H_0$ is the usual time-independent Hamiltonian of the unperturbed harmonic oscillator. Then, in interaction picture, we have $U = T e^{-i \lambda \int^t_0 H^I_{\rm int}(t') dt'}$. In the Magnus expansion, the terms in the Dyson series are collected and expressed as an exponential:
\begin{equation}
U = e^{-i \sum \limits_{n=1} \lambda^n \int^t_0 \Omega_n(t') dt'}\,,
\end{equation}
where
\begin{equation}
\Omega_1 = H^I_{\rm int}\,, \qquad \Omega_2 = \frac{1}{2} \int^t_0 [H^I_{\rm int}(t'),H^I_{\rm int}(t'')] dt'' 
\end{equation}
and
\begin{equation}
\Omega_3 = \frac{1}{6} \int^t_0 \int^{t'}_0 ([H^I_{\rm int}(t'),[H^I_{\rm int}(t''),H^I_{\rm int}(t''')] 
+[H^I_{\rm int}(t'''),[H^I_{\rm int}(t''),H^I_{\rm int}(t')] ) dt'' dt''' ...
\end{equation}
The existence of the small parameter $\lambda$ enables us to approximate $U$ by truncating the expansion. The result is a single exponential, without time-ordering. But we cannot immediately read off the squeezing parameters because $U \sim SR$, which is a product of exponentials.

To proceed, note that commutators of quadratic ladder operators are themselves quadratic, and thus the quadratic operators obey an algebra, namely $su(1,1)$. Concretely, defining $L_+ =\frac{1}{2} (a^\dagger)^2$, $L_- = \frac{1}{2} a^2$ and $L_0 = \frac{1}{2}(a^\dagger a + \frac{1}{2})$ we find 
\begin{equation}
[L_0, L_\pm] = \pm L_\pm\,, \quad [L_+, L_-] = - 2 L_0\,.
\end{equation}
Now the Baker-Campbell-Hausdorff formula is representation independent. Let us choose a standard representation of $su(1, 1)$:
\begin{equation}
    \label{exprep}
L_{+}=\begin{pmatrix}0&0\\1&0\end{pmatrix}\,,\qquad
L_{0}=- \frac12\begin{pmatrix}1&0\\0&-1\end{pmatrix}\,,\qquad
L_{-}=\begin{pmatrix}0&-1\\0&0\end{pmatrix}\,.
\end{equation}
Then the product of the squeezing and rotation operators is
\begin{equation}
\label{SRrepn}
S R 
    = e^{r( e^{- i \varphi} L_- - e^{i \varphi} L_+)}  e^{-i \theta  (2 L_0 -1/2)} = 
    e^{-i \theta/2}
    \left(
\begin{array}{cc}
 e^{2 i \theta } \cosh r & -e^{-i
   \varphi } \sinh r \\
 -e^{i (2 \theta +\varphi )} \sinh r
   & \cosh r\\
\end{array}
\right)\,.
\end{equation}
Meanwhile the result of the Magnus expansion is also an exponential of quadratic ladder operators. Hence it can be written as
$U = \exp(\zeta) \exp( c_+ L_+ + c_- L_- + c_0 L_0)$, where $\zeta$ is a c-number phase and the coefficients $c_{\pm}, c_0$ are computable as a power series in $\lambda$. (By unitarity, $c_- = -c_+^*$ and $c_0$ is imaginary.) Then, using the representation \eqref{exprep}, we have
\begin{equation}
\label{Magnusrepn}
\exp(\zeta) 
\exp( c_+ L_+ + c_- L_- + c_0 L_0) = 
\exp(\zeta) 
\left(
\begin{array}{cc}
 \cosh (b )-\frac{c_0 \sinh (b)}{2
   b} & -\frac{c_-  \sinh (b)}{b} \\
 \frac{c_+  \sinh (b)}{b } &
   \frac{c_0  \sinh (b )}{2 b }+\cosh
   (b ) \\
\end{array}
\right)
\end{equation}
where $b = \frac{1}{2} \sqrt{4 |c_+|^2 -|c_0|^2}$. We can now read off the squeezing parameters by equating the matrices in \eqref{SRrepn} and \eqref{Magnusrepn}. In particular, one readily finds $\zeta = \frac{i \theta}{2}$ and thus $r =  \, \mathrm{arcsinh}(|\frac{c_+ \sinh b}{b}|)$.

\section{Scalar Field Theory}
Before turning to gravity, let us briefly examine scalar field theory to demonstrate the complications arising in field theory in a somewhat simpler setting. Consider, then, a real scalar field with a time-dependent quadratic potential on Minkowski space
\begin{align}
        S &= \frac{1}{2}
        \int_{\mathbb{R}} dt
        \int_V d^3x \,  \left(\dot{\phi}^2 - {(\nabla \phi)}^2 + f(t) \phi^2\right)
\end{align}
in a finite spatial volume $V$.
Quantizing the field in discrete modes
\begin{equation}
\label{fieldQuant}
    \phi = \frac{1}{\sqrt{V}} \, \sum_{\vec k} \frac{1}{\sqrt{ 2 \omega_{ k}}} \left(a_{\vec k} \, e^{i \cdot \vec{k} \vec{x}} + a^{\dagger}_{{\vec k}} \, e^{-i \vec{{\vec k}}  \cdot \vec{x}}\right)\,,
\end{equation}
where $\omega_k = \omega_{|\vec k|}$,s
leads to the Hamiltonian
\begin{align}
    \label{2modeHamiltonianf(t)}
        H = \sum_{\vec k}  \omega_{ k} \left( a^{\dagger}_{{\vec k}} a_{{\vec k}} + \frac{1}{2}\right) -  \frac{f(t)}{4\omega_{{ k}}}  \left( a_{{\vec k}} a_{-{\vec k}} + a^{\dagger}_{{\vec k}} a^{\dagger}_{-{\vec k}} + a^{\dagger}_{{\vec k}} a_{{\vec k}} + a_{{\vec k}} a^{\dagger}_{{\vec k}}\right)\,,
\end{align}
where we have used the normalization condition in finite volume $
    \int d^3 x \, e^{i(\vec{k}-\vec{k'}) \cdot \vec{x}} = V \, \delta_{\vec{k},\vec{k'}}
$.
The modes ${\vec k}$ and $-{\vec k}$ can be decoupled by defining two new sets of creation and annihilation operators
\begin{equation}
        b_{{\vec k}} \, {e^{-i \frac{\pi}{4}}} = \frac{(a_{\vec k} - i \, a_{-{\vec k}})}{\sqrt{2}} \, , ~~~~~ c_{{\vec k}} \, {e^{-i \frac{\pi}{4}}} = \frac{(a_{-{\vec k}} - i \, a_{{\vec k}})}{\sqrt{2}}\,,
\end{equation}
so that the Hamiltonian reads
\begin{equation} \label{QFT-Hjusttime}
    \begin{split}
        H
        &=  \sum_{\vec k}  \frac{\omega_k}{{2}} \left( b^{\dagger}_{{\vec k}} b_{{\vec k}}
        + c^{\dagger}_{{\vec k}} c_{{\vec k}}
        + {1}\right) - \frac{f(t)}{8 \omega_{k}} \,  \, \left( b_{\vec k}^{\dagger} b_{\vec k} + b_{\vec k} b_{\vec k}^{\dagger} {+} b_{\vec k}^2 + {b^{\dagger}_{\vec k}}^2 + c_{\vec k}^{\dagger} c_{\vec k} + c_{\vec k} c_{\vec k}^{\dagger} {+} c_{\vec k}^2 {+} {c^{\dagger}_{\vec k}}^2\right)\,.
    \end{split}
\end{equation}
Thus, the problem reduces to two decoupled harmonic oscillators in a time-dependent quadratic potential for each $k$. Comparison with \eqref{hamiltonian} yields 
\begin{equation}
F(t) = \frac{f(t)}{{8}} \,, \qquad 
\omega_0 = \frac{\omega_k}{{2}} \,.
\end{equation}
The squeezing parameter for each mode can be calculated via the methods of Section \ref{sec:harm}.

Next, instead of considering a time-dependent potential, we consider a spacetime dependent potential of the form $f(t, \vec{x}) \, \phi^2$, which contributes to the Hamiltonian as
\begin{align}
\label{QFT-IntHamiltonian}
       - \frac{1}{4}\sum_{{\vec k},{\vec k}^\prime} \frac{1}{\sqrt{\omega_{ k} \omega_{{ k}'}}} 
       \left(
       f_{-{\vec k}-{\vec k}'}(t) a_{{\vec k}} a_{{\vec k}'} + f_{{\vec k}+{\vec k}'}(t) a^{\dagger}_{{\vec k}} a^{\dagger}_{{\vec k}'}
       + f_{{\vec k}-{\vec k}'}(t) a^{\dagger}_{{\vec k}} a_{{\vec k}'} + f_{-{\vec k}+{\vec k}'}(t) a_{{\vec k}} a^{\dagger}_{{\vec k}'}
       \right)
\end{align}
where $f(x,t) = \sum_{\vec k} \, f_{\vec k}(t) e^{i {{\vec k}} \cdot \vec{x}}$ with $f^*_{{\vec k}}=f_{-{\vec k}}$.
Unlike \eqref{QFT-Hjusttime}, the Hamiltonian does not reduce to a sum of decoupled harmonic oscillators. In general, this makes extracting the squeezing parameters much more difficult, in each of our techniques. The numerical evolution now involves an infinite number of nonlinear coupled differential equations, instead of just two; the Bogolyubov method now requires diagonalizing infinite-dimensional matrices; the perturbative method now involves $sp(2N,\mathbb{R})$ matrices for infinite $N$ in lieu of $su(1,1)$. Fortunately, the existence of compact symmetries (such as rotational symmetry) makes the problem tractable again, as the modes then split up into finite-dimensional representations. 

Consider, for example, a finite number of modes $\vec{k} \in \{ k_1,k_2,...,k_N \} \, , k_i \in \mathbb{Z}^3$ that mix among themselves and for which the Bogolyubov coefficients are known. Then the squeezing parameters can be extracted as follows. The time evolution operator, $U$, associated to the Hamiltonian obeys
\begin{equation}
\label{Bogrelations}
        U^{\dagger} \, \begin{pmatrix}
            a \\ a^{\dagger}
        \end{pmatrix} \, U = \begin{pmatrix}
            \alpha^* && -\beta^* \\ -\beta && \alpha
        \end{pmatrix} \,  \begin{pmatrix}
            a \\ a^{\dagger}
        \end{pmatrix}
\end{equation}
where $a \equiv \left( a_{k_1} \, , a_{k_2} \, , .... \right)^T$ and $a^{\dagger} \equiv ( a_{k_1}^{\dagger} \, , a_{k_2}^{\dagger} \, , .... )^T$ and $\alpha$, $\beta$ are the Bogolyubov matrices. These matrices describe the transformation between old and new annihilation and creation operators in the Heisenberg picture. They satisfy
\begin{subequations}
\begin{align}
        \sum_{\vec{k}} \, \alpha_{\vec{i} \, \vec{k}} \, \alpha^*_{\vec{j} \, \vec{k}} - \beta_{\vec{i} \, \vec{k}} \, \beta^*_{\vec{j} \, \vec{k}} &= \delta_{\vec{i} \, \vec{j}} \, , \label{Bogconstraints1} \\
        \sum_{\vec{k}} \, \alpha_{\vec{i} \, \vec{k}} \, \beta_{\vec{j} \, \vec{k}} - \beta_{\vec{i} \, \vec{k}} \, \alpha_{\vec{j} \, \vec{k}} &= 0\,. \label{Bogconstraints2}
\end{align}
\end{subequations}
Incidentally, the number of independent Bogolyubov coefficients from \eqref{Bogrelations} is $N(2 N+1)$ ($4 N^2$ minus $N^2$ constraints from \eqref{Bogconstraints1} minus $N(N-1)$ constraints from \eqref{Bogconstraints2}), which matches precisely the number of independent parameters in the most general quadratic Hamiltonian \footnote{The most general quadratic Hamiltonian reads
    $H = \sum_{i,j} \xi_{k_i k_j} \, a^{\dagger}_{k_i} \, a_{k_j} + \gamma_{k_i k_sj} \, a_{k_i} \, a_{k_j}^{\dagger} + \Gamma^*_{k_i k_j} \, a^{\dagger}_{k_i} \, a^{\dagger}_{k_j} + \Gamma_{k_i k_j} \, a_{k_i} \, a_{k_j} $
where, by hermiticity, one has
$\xi = \xi^*$ ($N^2$ real parameters) ,  
        $\gamma = \gamma^*$ ($N^2$ real parameters) and 
        $\Gamma = \Gamma^T$ ($N(N+1)$ real parameters).
The canonical commutation relations $[ a_i,a^{\dagger}_j] = \delta_{ij}$ impose $N^2$ conditions, yielding $N^2 + N^2 + N(N+1) - N^2 = N(2N+1)$ independent parameters.}.

The time evolution operator
$U$ can be written as a product of a multi-mode squeezing operator and a multi-mode mode rotation operator multiplied by an overall phase  (cf.\ \cite{PhysRevA.94.062109, Ma:1990llj}) 
\begin{equation}
    \begin{split}
        U = e^{i \delta} \, S(r, \varphi) \, R(\theta)\,,
    \end{split}
\end{equation}
where all quantities depend on time; $\varphi,\theta$ are Hermitian matrices, $r$ is a positive semidefinite matrix 
and $\delta$ is a real number
\begin{align}
\label{multimode}
        S(r, \varphi) &= \exp \bigg( \frac{1}{2} \, \sum_{\vec{k},\vec{k}'} \, \left( (e^{-i \varphi} \,r  )_{\vec{k},\vec{k}'} \, a_{\vec{k}} \, a_{\vec{k}'} - (r e^{i \varphi})_{\vec{k},\vec{k}'} \, a^{\dagger}_{\vec{k}} \, a^{\dagger}_{\vec{k}'} \right) \bigg) \,,  \\
        R(\theta) &= \exp \bigg( -i \sum_{\vec{k},\vec{k}'} \, \theta_{\vec{k},\vec{k}'} \, a^{\dagger}_{\vec{k}} \, a_{\vec{k}'} \bigg)\,.
\end{align}

The action of these operators on the column vectors are given as 
\begin{subequations}
    \begin{align}
        S^{\dagger} \, a \, S &= \cosh{r} \, a - \sinh{r} \, e^{i \varphi} \, a^{\dagger} \, ,   &S^{\dagger} \, a^{\dagger} \, S &= \cosh{r^T} \, a^{\dagger} - \sinh{r^T} \, e^{-i \varphi^T} \, a \\
        R^{\dagger} \, a \, R &= e^{-i \theta} \, a \, , & R^{\dagger} \, a^{\dagger} \, R  &= e^{-i \theta^T} \, a^{\dagger}
    \end{align}
\end{subequations}
Here, the displacement operator has already been omitted for the same reason as in section \ref{sec:timeevolution}.

When evolved from vacuum, the rotation operator acts trivially, and the state of the system is now a multi-mode squeezed state:
\begin{equation}
|\Psi  \rangle = \exp \bigg ( \frac{1}{2} \sum_{\vec{k},\vec{k}'}(e^{-i \varphi} r)_{\vec{k}, \vec{k}'}  a_{\vec{k}} a_{\vec{k}'} - (r e^{i \varphi})_{\vec{k}, \vec{k}'} a^\dagger_{\vec{k}} a^\dagger_{\vec{k}'} \bigg )|0 \rangle
\end{equation}
where $r_{\vec{k},\vec{k}'}$ is now a matrix of squeezing parameters, which can be obtained provided the Bogolyubov coefficients are known. In that case,
\begin{equation}
\label{matrixeqsinhr}
    \sinh{r} \, e^{i \varphi} \, e^{i \theta} = \beta^*  
\end{equation}
which generalizes \eqref{rbog}.
Multiplying \eqref{matrixeqsinhr} by its adjoint yields
    $(\sinh{r})^2  = \beta^* \beta^T$.
Since $\sinh(r)$ is positive semidefinite (as $r$ is positive semidefinite), it can be brought into diagonal form with eigenvalues which are greater or equal to zero. In this way, the squeezing parameters can be read off.

Perhaps the simplest way to extract the multi-mode squeezing parameters in field theory is using the Magnus expansion method. Recall that the interaction Hamiltonian typically contains terms of the form $\theta_{\vec{k}\vec{k}'} a^\dagger_{\vec{k}} a_{\vec{k}'}$, which contribute not only to the rotation operator $R$, but also to the squeezing operator $S$; this can be seen even from \eqref{schum}, wherein $\theta$ evidently contributes to the dynamics of $r$. In the case of a single oscillator, we were able to use a representation of $su(1,1)$ to read off the squeezing parameter. In the most general field theory case, however, we would have to formally work with representations of $sp(2N,\mathbb{R})$ in the infinite $N$ limit. But fortunately, to leading order (in the small parameter $\lambda$ that organizes the Magnus expansion), we can neglect all commutators:
\begin{align}
U 
 &= T e^{-i \lambda \int^t_0 H^I_{\rm int}(t') dt'}  \nonumber\\
&\sim e^{-i \lambda \int^t_0 H^I_{\rm int}(t') dt' + {\cal O}(\lambda^2)} \nonumber \\
&\sim e^{\frac{i}{4}\lambda \sum_{\vec{k}, \vec{k}'}
\frac{1}{\sqrt{\omega_{ k} \omega_{{ k}'}}}
\int^t_0 \left ( f_{\vec{k} \vec{k}'} a^{I, \dagger}_{\vec{k}} a^{I, \dagger}_{\vec{k}'} dt' +h.c. \right ) dt'}e^{-i \lambda \sum_{\vec{k}, \vec{k}'} \int^t_0 \left(\theta_{k,k'}(t') a^{I, \dagger}_{\vec{k}} a^I_{\vec{k}'} + h.c. \right) dt' }\,.
\label{ufield}
\end{align}
The second exponential gives a pure phase when acting on the vacuum. Comparing with the definition of the multi-mode squeezing operator \eqref{multimode}, we find
\begin{equation}
 (r \,e^{i \varphi} )_{\vec{k},\vec{k}'}=  - \frac{i}{2} \frac{1}{\sqrt{\omega_{ k} \omega_{{ k}'}}} \int^t_0 f_{\vec{k} \vec{k'}}(t') e^{ i (\omega_{\vec{k}} + \omega_{\vec{k}'}) t^\prime} dt' \,.
 \label{rfield}
\end{equation}
In this approximation, the squeezing parameters can be read off from the time integral of the Fourier coefficients of the interaction Hamiltonian.

\section{Gravity}
We have seen that time-dependent prefactors in the quadratic terms generically take vacuum states to squeezed states. Such terms do not naturally arise in a self-contained way in scalar field theory or electrodynamics. For example, the $f(t) \phi^2$ term in the scalar field action corresponds to a $J \phi^2$ coupling, rather than the usual $J \phi$ term. In electrodynamics, the form of the Lagrangian is such that time-dependent classical background fields, $A^{\mu}_c(t,x)$, do not couple to the perturbations; to obtain a time-dependent prefactor for the quadratic perturbations, one has to instead couple the gauge field to some other field, such as a time-dependent spacetime metric. By contrast, as we demonstrate, the standard Einstein-Hilbert action for gravity naturally contains terms that can lead to the production of squeezed states. This is because the inverse metric, the metric determinant, and the Riemann tensor all contain arbitrary powers of the metric perturbation. Thus expanding $\sqrt{-g}g^{ac}g^{bd}R_{abcd}$ about a time-dependent solution automatically results in time-dependent prefactors in front of the quadratic perturbations; the same is true for the matter action.

Let us start by considering the action of general relativity in four spacetime dimensions
\begin{equation}
    S[g, \Psi] = \frac{1}{2 \mykappa} 
    \int\,
    \sqrt{-g}\, d^4 x\,  (R-2 \Lambda)+ S_m[\Psi]\,,
\end{equation}
where $\kappa = 8 \pi G$
and we denote any matter fields as $\Psi$. 
The first variation of the action with respect to the metric 
\begin{equation}
\label{firstvariation}
   S^{(1)}[\bar g, h, \Psi]
   = -
   \frac{1}{2 \mykappa}
   \int \sqrt{-\bar{g}}\, d^4 x  \left( \bar{R}_{a b} - \frac{1}{2} \bar{R} \bar{g}_{a b}
   + \Lambda \bar{g}_{a b} - \mykappa T_{a b}(\bar{g}, \Psi)
   \right) h^{ab}\,,
\end{equation}
yields the equations of motion for $\bar{g}$.
Here, we vary with respect to the metric and $\delta_g g_{ab} = h_{ab}$ (and $\delta_g g^{ab} = -h^{ab}$).
The second variation is 
\begin{align}
   S^{(2)}[\bar g, h, \psi] &= -
     \frac{1}{8 \mykappa}\int
    \sqrt{-\bar g} \,d^4 x  h^{ab}\Big( 
   2 \Lambda h_{ab} -  \bar{R}  h_{ab}
    + \bar{g}_{ab}\bar{R}^{cd} h_{cd} - \bar{\nabla}_a \bar{\nabla}_b h^c_c 
    + {{\bar \nabla}_a {\bar \nabla}^c h_{cb} }
+
{{\bar \nabla}_b {\bar \nabla}^c h_{ca} }
    \nonumber \\ 
      &~~~  
      +  \bar{R}_{a}^{~c} h_{cb}
+ \bar{R}_{b}^{~c} h_{ca}
    - 2 \bar{R}_{facb}  h^{fc}
      -  \bar{\Box} h_{ab}
      -  \bar{g}_{ab} \bar{\nabla}^c \bar{\nabla}^d h_{cd} +  \bar{g}_{ab}
      \bar{\Box} h^c_c - 2 \mykappa \, \delta_g T_{ab}
    \Big) \,.\label{secondvarGRgeneral1}
\end{align}
where $\delta_g T_{ab}$ denotes the variation of the energy-momentum tensor with respect to the metric and
where we have used that the background $\bar{g}$ satisfies the equations of motion.

One can already see that there are a number of terms that are quadratic in the metric perturbation and are coupled to curvature tensors. Time-dependence of such tensors can therefore supply the necessary ingredient for squeezing. So far, we have not fixed any gauge.

Under linearized diffeomorphism $\xi$, the metric transforms as 
$
    h_{ab}
    \rightarrow h_{ab } -
    \bar{\nabla}_a \xi_b -
    \bar{\nabla}_b \xi_a\,.
$
Define the trace-reversed perturbation as usual
\begin{equation}
    \tilde{h}_{a b} = h_{a b} - \frac{1}{2} h^c_{~c} \bar{g}_{a b}\,.
\end{equation}
This transforms as 
$
   \tilde{h}_{a b} \rightarrow \tilde{h}_{a b} - \bar{\nabla}_a \xi_b -
    \bar{\nabla}_b \xi_a + 
    \bar{\nabla}^a \xi_a\,
    \bar{g}_{ab}$.
Thus, under a linearized diffeomorphism
  we have 
\begin{equation}
 \bar{\nabla}^{a} \tilde{h}_{a b}\rightarrow \bar{\nabla}^{a} \tilde{h}_{a b} - \bar{\Box} \xi_a - \bar{R}_{c a} \xi^c\,.
\label{lindiffeo}
\end{equation}
Unique solutions to the equation $\bar{\nabla}^{a} \tilde{h}_{a b} - \bar{\Box} \xi_a - \bar{R}_{c a} \xi^c = 0$ always exist on globally hyperbolic spacetimes, cf.\ e.g. 
 \cite[Thm. 10.1.2]{Wald:1984rg}.
 Hence, the harmonic gauge 
 \begin{equation}
 \bar{\nabla}^{a} \tilde{h}_{a b} = 0
 \end{equation}
 is always achievable (on globally hyperbolic spacetimes). In this gauge, \eqref{secondvarGRgeneral1} reduces to 
 \begin{align}
   S^{(2)}[\bar g, h] &= -
     \frac{1}{8 \mykappa}\int
    \sqrt{-\bar g} \,d^4 x  h^{ab}\Big( 
   2 \Lambda h_{ab} -  \bar{R}  h_{ab}
    + \bar{g}_{ab}\bar{R}^{cd} h_{cd} 
    +  
      \bar{R}_{a}^{~c} h_{cb}
+ \bar{R}_{b}^{~c} h_{ca} - 2 \bar{R}_{facb}  h^{fc}
     \nonumber \\ 
      &~~~
      -  \bar{\Box} h_{ab}
     +  \frac{1}{2}\bar{g}_{ab}
      \bar{\Box} h^c_c 
      - 2\mykappa \,\delta_g T_{ab}
    \Big) \,.
    \label{genaction2}
\end{align}
However, harmonic gauge does not completely fix the gauge freedom; usually, on Ricci-flat backgrounds the tracelessness condition $h^c_c = 0$ is also imposed. But a necessary condition for this is that $\bar{\Box} h^c_c = 0$, which is met if the background and the perturbed spacetime \emph{both} satisfy the vacuum equations, which is generally not the case for off-shell quantum perturbations. To see that $\bar{\Box} h^c_c = 0$ is necessary, note that, under linearized diffeomorphisms, $h^c_c$ transforms as $h^c_c \rightarrow h^c_c - 2  \bar{\nabla}^c \xi_c$. Hence we want to find a residual diffeomorphisms such that $h^c_c - 2  \bar{\nabla}^c \xi_c= 0$. As $\bar{\Box}  h^c_c - 2 \bar{\Box}  \bar{\nabla}^c \xi_c = \bar{\Box} h^c_c - 2  \bar{\nabla}^c \bar{\Box} \xi_c = \bar{\Box} h^c_c$ (for $\bar{R}_{a b} = 0$) for these residual diffeomorphisms, a necessary condition to impose $h^c_c = 0$ in addition to $\bar{\nabla}^{a} \tilde{h}_{a b} = 0$, is that $\bar{\Box} h^c_c = 0$. Since a fully gauge-fixed off-shell action is not available, the tracelessness condition has to be dealt with by constraining the physical Hilbert space. 

Let us now consider various examples illustrating our formalism to demonstrate how squeezed states arise.

\subsection{FRW}
Consider first the squeezing of gravitational perturbations on a Friedmann–Robertson–Walker (FRW) background. This question was already studied in \cite{Grishchuk:1975uf, Grishchuk:1989ss, Grishchuk:1990bj}, as well as more recently in \cite{Kanno:2021vwu}; this section essentially reviews the results of \cite{Kanno:2021vwu} in connection to the present work. 

We assume that the variation of the energy-momentum tensor of the matter satisfies \begin{equation}
    \delta_g T^a_{~b} = 0\qquad
    \Rightarrow  \qquad 
    \delta_g T_{ab} = \frac{1}{2}
   (h_{ac} \bar{T}^c_{~b} + h_{bc} \bar{T}^c_{~a})\,,
\end{equation}
an assumption valid, for example, in the case of a minimally coupled scalar field.
Under this assumption \eqref{genaction2} reduces to 
 \begin{align}
   S^{(2)}[\bar g, h] &= -
     \frac{1}{8 \mykappa}\int
    \sqrt{-\bar g} \,d^4 x  h^{ab}\Big( 
    \bar{g}_{ab}\bar{R}^{cd} h_{cd} - \bar{\nabla}_a \bar{\nabla}_b h^c_c  
    - 2 \bar{R}_{facb}  h^{fc}
      -  \bar{\Box} h_{ab}
     +  \frac{1}{2}\bar{g}_{ab}
      \bar{\Box} h^c_c 
    \Big) \,
\end{align}
where the equations of motion for the background $\bar{g}$ have been used.
We now consider perturbations of FRW such that the metric in conformal coordinates reads
\begin{equation}
 ds^2 =  (\bar{g}_{ab} + h_{ab}) dx^a dx^b = 
     (a^2(\eta) \eta_{a b} + h_{a b}) dx^a dx^b\,.
\end{equation}
It is well known, see e.g.\ \cite{Maggiore:2018sht}, that such perturbations can be decomposed into scalar, vector, and tensor components according to their transformation properties under spatial rotations. The linearized equations of motion and hence the second variation of the action fully decouple into scalar, vector, and tensor sectors when evaluated on the FRW background. Among these, only the tensor perturbations correspond to propagating gravitational waves in the classical theory.
For this reason, and to keep the analysis as simple as possible, we restrict ourselves to tensorial fluctuations, which satisfy
\begin{equation}
\label{transversetraceless}
h_{00} = h_{0 i} = 0\,, \qquad
  \delta_{i j} h^{i j} = 0\,, \qquad
  {\bar \nabla}^j h_{i j} =  \partial^j h_{i j} = 0\,.
\end{equation}
The second variation of the action \eqref{secondvarGRgeneral1} then reduces to
\begin{align}
\label{expansionfull}
    S^{(2)}[\bar g, h] &=
    \frac{1}{8 \mykappa}\int
    \sqrt{-\bar g} \,d^4 x \, h^{ij}
    \Big( 
    \bar{\Box} h_{ij} +
     2 \bar{R}_{kilj}  h^{kl}
    \Big) \,.
\end{align}
Upon expanding the field $h_{ij}$ in Fourier modes as 
\begin{equation}
\label{modesfourier}
    h_{i j}(\eta, x) = \sum_{s, \vec k}
    q_{\vec k, s}(\eta) \epsilon_{i j}^{(s)} (\vec k) e^{i \vec{k} \vec{x}}\,,
\end{equation}
where $q_{\vec k, s}$ is subject to
\begin{equation}
\label{qindetail}
q^*_{\vec k, s} \epsilon_{ij}^{(s)}(\vec k) = q_{-\vec k, s}
    \epsilon_{ij}^{(s)}(-\vec k)\,, \qquad 
    \epsilon_{ij}^{(s)} (\vec k) \epsilon_{ij}^{(s^\prime)}(\vec k) = 2 \delta^{s, s^\prime}\,,
\end{equation}
decomposing as 
\begin{equation}
\label{decompose}
q_{\vec k s}(\eta) = 
\frac{a(\eta)}{l_P } X_{\vec k s}(\eta)  
+ i \frac{a(\eta)}{l_P } Y_{\vec k s}(\eta)   
\,,
\end{equation}
where $l_P$ is the Planck length,s
and inserting the above into \eqref{expansionfull}, we obtain
\begin{align}
  S^{(2)}[\bar g, h]
&= \frac{V}{2 \mykappa l_P^2} 
 \int d \eta 
 \sum_{\vec k, s}
 \Big \{\frac{1}{2}
  (\dot X_{\vec k s}(\eta)^2
+ \dot Y_{\vec k s}(\eta)^2) 
+ \frac{1}{2}
\Big( \frac{a''(\eta )}{a(\eta )} - \vec k^2 
   \Big)(X_{\vec k s}(\eta)^2
+ Y_{\vec k s}(\eta)^2) 
 \Big \}\,.
\end{align}
Here, we have assumed that the field $h_{ab}$ falls off at early and late  times and we have used the orthonormality condition in finite volume  $
    \int d^3 x e^{i (\vec k - \vec k^\prime)\cdot  \vec{x}} = V \delta_{\vec k, \vec k^\prime}$.

Thus, the problem reduces to two decoupled harmonic oscillators in a time-dependent quadratic potential for each $s$ and $k$. Comparison with \eqref{eq:Lagrangianharm}, yields 
\begin{equation}
F(\eta) = \frac{1}{2} \frac{a''(\eta )}{a(\eta )}\,, \qquad 
\omega_0 = \sqrt{\vec k^2}
\end{equation}
As an example, we consider de Sitter space, where $F(\eta) = 1/\eta^2 $, and thus,
for each mode and polarization, the classical equation of motion reads
\begin{equation}
    \begin{split}
        {X''} + k^2 \, X - \frac{2}{\eta^2} X = 0\,.
    \end{split}
\end{equation}
Solving this equation subject to the initial conditions $X(\eta_0) = \frac{1}{\sqrt{2 \, \omega_0}}, X'(\eta_0) = -i \, \omega_0 \, X(\eta_0)$, we can infer the squeezing parameter for the $k$th mode by
using \eqref{alpha-beta} and \eqref{rbog}:
\begin{align}
   \sinh( r_k(\eta) ) &= \frac{ \big(1+2 \left( \eta^4 + \eta_0^4 \right)\, \omega_0^4 + A(\eta) \, \sin{[2 \left( \eta-\eta_0 \right) \, \omega_0]} + B(\eta) \, \cos{[2 \left( \eta-\eta_0 \right) \, \omega_0]} \big)^{\frac{1}{2}}}{2 \sqrt{2} \, \eta_0^2 \, \omega_0^4 \, \eta^2}\,,
\end{align}
where $\omega_0 = \sqrt{k^2}$,
$A(\eta) = -2 \left( \eta-\eta_0 \right) \, \omega_0 \, \left( 1+2 \eta \, \eta_0 \, \omega_0^2 \right)$ 
and $B(\eta) = -1+2 \, \left( \eta-\eta_0 \right)^2 \, \omega_0^2 -4 \eta^2 \eta_0^2 \, \omega_0^4$.
The behavior as $\eta \rightarrow 0^-$ is given by 
\begin{equation}
   r_k(\eta) 
   \sim C - 2 \mathrm{log}\left(\omega_0 |\eta |\right)\,,
\end{equation}
where $C$ is constant. The asymptotic form of $r_k$ exactly matches that found by Grishuk \cite{Grishchuk:1975uf}. For de Sitter space, one finds that
\begin{equation}
\frac{dr_k}{dt} = H
\end{equation}
so that the squeezing parameters for all modes increase by one after each Hubble time.

\subsection{Kasner}
As another example, consider the Kasner spacetime, a special case of the homogeneous but anisotropic Bianchi Type-I universe that is often used to model pre-inflationary cosmology. Unlike the FRW examples considered earlier, the Kasner metric is a vacuum spacetime so the time-dependent potential originates in the Riemann tensor that comes from commuting covariant derivatives of the metric perturbations. The line element can be written as
\begin{equation}
   ds^2 = - dt^2 + \sum_{j =1}^3 t^{2 p_j} (dx^j)^2\,,
\end{equation}
where $p_j$ must satisfy
\begin{equation}
   \sum_{j = 1}^3 p_j = 1\,, \qquad    \sum_{j = 1}^3 p_j^2 = 1
\end{equation}
to comprise an exact solution of the vacuum Einstein equations with vanishing cosmological constant. 

\begin{figure}
    \centering
 \includegraphics[width=0.47\linewidth]{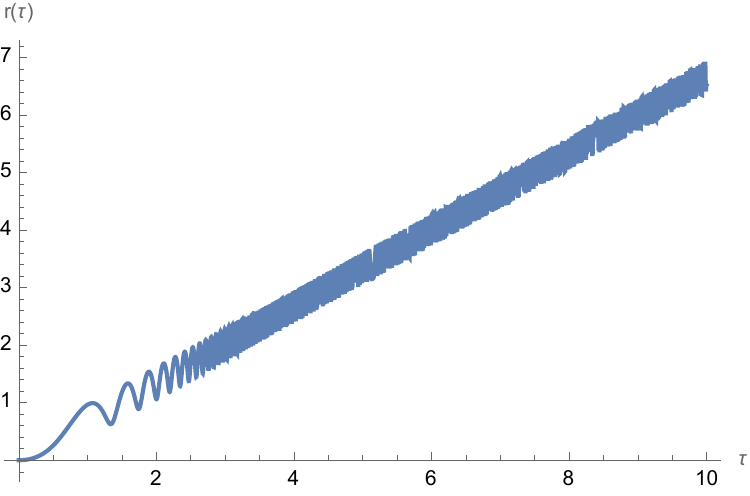}
    \caption{Plot of the squeezing parameter $r(t)$ for the decoupled perturbation modes in a Kasner universe. Note the linear growth of the squeezing parameter with time.}
    \label{fig:placeholder}
\end{figure}

We now consider the special case of $p_1= -\frac{1}{3}, p_2=p_3=\frac{2}{3}$ and for simplicity choose a different gauge, radiation gauge $h_{0 \mu} = 0$. Furthermore, we restrict to classes of perturbations which only depend on one spatial coordinate, $h = h(t, x^1)$. It was shown in \cite{PhysRevD.18.969} that for such perturbations,
upon computation of the linearized equations of motions, the $h_{23}$-component, as well as the combination $h_{22}-h_{33}$
completely decouple and satisfy a scalar wave equation,  
with wave vector $\vec k= \{k_1, 0, 0 \} $.
Indeed, \cite[see (2.8)]{PhysRevD.18.969}
\begin{equation}
\label{eqkasner23}
    \frac{d^2}{dt^2}\phi_{(A)} + \kappa  \frac{d}{dt}\phi_{(A)} + \omega^2 \phi_{(A)} = 0\,,
\end{equation}
where $\phi_{(1)} = \frac{{\tilde h}_{23}}{a_2 a_3}$ or $\phi_{(2)} = \frac{1}{2} \left(\frac{{\tilde h}_{22}}{a_2^2} - \frac{{\tilde h}_{33}}{a_3^2} \right)$, $a_j = t^{p_j}$ 
with $\kappa = \sum_j \frac{\dot a_j}{a_j}$ and $\omega = \frac{k_1}{a_1}$. 
By defining a new temporal coordinate
$
     \tau = \ln(t)
$
we can write \eqref{eqkasner23} as
\begin{equation}
\label{eqkasner24}
   \frac{d^2}{d \tau^2} \phi_{(A)} + e^{\frac{8 \tau}{3}} \, k_1^2 \, \phi_{(A)} = 0\,.
\end{equation}
This is again the equation of motion of a harmonic oscillator in a time-dependent potential.
With $k_1=1$ and $\phi_{(A)}'(\tau=0)=-i \phi_{(A)}(\tau=0)$ and $\phi_{(A)}(\tau=0) = 1/\sqrt{2}$, the solution for our mode reads
\begin{equation}
    \frac{8 \sqrt{2}}{3 \pi}\phi_{(A)} (\tau) =  i  \left (iY_1 \left (\frac{3}{4} \right ) + Y_0 \left (\frac{3}{4}\right)\right) \, _0\tilde{F}_1\left(1;- \frac{9}{64}  e^{\frac{8 \tau }{3}}\right)+\left(-i J_0\left(\frac{3}{4}\right)+J_1\left(\frac{3}{4}\right)\right) Y_0\left(\frac{3}{4} \sqrt{e^{\frac{8 \tau }{3}}}\right) \,,
\end{equation}
where $J_\alpha(x)$ and $Y_\alpha(x)$ are Bessel functions and ${}_p\tilde{F}_q$
is the regularized hypergeometric function. The mode solution can be inserted into \eqref{alpha-beta} and \eqref{rbog} to obtain the squeezing parameter; this is plotted in Figure \ref{fig:placeholder}.

\subsection{Squeezing from Matter}
So far we have considered examples where squeezing occurs due to time-dependent background curvature. Now we consider the effect of time-dependent sources. As we have seen repeatedly, the necessary condition for production of squeezed states is a time-dependent coefficient of the quadratic term in the field. For a scalar field, one therefore would need a quadratic coupling of the source of the form $J(t) \varphi^2$. By contrast, in electrodynamics, such a term is forbidden by gauge invariance; thus, time-dependent classical charges and currents in electrodynamics generically produce coherent states, rather than squeezed states. 

For gravity, squeezed states are produced by time-dependent sources when the second variation of the matter action with respect to the metric is non-vanishing. Consider the example of a point particle in the absence of non-gravitational forces:
\begin{equation}
\label{Spp}
    S_M[g, X] = - 
m \int d\tau
    \sqrt{-g_{a  b} \dot{X}^a \dot{X}^b}\,.
\end{equation}
Here $X^a  (\tau)$ denotes the position vector of the particle in spacetime in contrast to $x^a $ which denotes the coordinates; $\tau$ is an arbitrary parameter.  
The first variation of the action reads
\begin{equation}
     S_M^{(1)}[\bar g, h, X]
    = \frac{m}{2} 
    \int d\tau
    \, 
    \frac{ h_{a  b}
    \dot{X}^a \dot{X}^b}{\sqrt{-\bar{g}_{c d} \dot{X}^c  \dot{X}^d}}\,.
\end{equation}
One could write this as a spacetime integral by inserting $\int d^3 x\, \delta^{3}(x-X)$ to obtain the energy-momentum tensor, but we will not need that here. The second variation of the action reads
\begin{equation}
\label{varact2}
     S_M^{(2)}[\bar g, h, X]
    =    \frac{m}{8} 
    \int  d\tau \,
    \frac{h_{a  b}
    \dot{X}^a \dot{X}^b
     h_{c d}
    \dot{X}^c \dot{X}^d}{(-\bar{g}_{e  f} \dot{X}^e \dot{X}^f)^{3/2}} \,.
\end{equation}

Let us now focus on a concrete example. Consider a massive particle in orbit around a Schwarzschild black hole of mass $M$; this could represent a binary system with an extreme mass ratio, $m \ll M$, such as the (unconfirmed) neutron star-black hole merger event GW191219 \cite{KAGRA:2021vkt}, estimated to have a mass ratio of 0.09. Now, when the particle is sufficiently far away, the Schwarzschild geometry can be approximated as  flat spacetime with a weak Newtonian potential, $\Phi$. The line element is
\begin{equation}
        \bar{g}_{a b} dx^a dx^b= -\left( 1+2 \Phi \right) \, dt^2 + \left( 1-2 \Phi\right) \, \left( dx^2 + dy^2 + dz^2 \right)\,,
        \label{lineelem0}
        \end{equation}
where $\Phi = - \frac{\kappa M}{8 \pi r}$. Since the geometry is static, there will be no contribution to the squeezing from the Einstein-Hilbert part of the total action. The generation of squeezed states by a bounded two-body system was studied by Kanno et al. \cite{Kanno:2025how}. They considered particles of arbitrary mass ratios undergoing circular orbit in flat space, whereas our starting point is the Schwarzschild geometry valid for extreme mass ratios. Formally, our results are broadly similar to theirs, up to a numerical factor; however, we isolate the dominant squeezing parameters in somewhat more detail. But crucially we find that, for certain modes, the squeezing parameter grows linearly in time, just as one might anticipate from our earlier analysis of the harmonic oscillator.

Incidentally, inserting \eqref{lineelem0} this into the matter action \eqref{Spp} yields
\begin{equation}
\label{83}
S_M = - m \int d \tau \sqrt{-\eta_{ab} \,\dot{X}^a \, \dot{X}^b} \left ( 1 + 
\frac{\Phi \,\delta_{cd}\, \dot{X}^c \, \dot{X}^d}{-\eta_{ef} \,\dot{X}^e \, \dot{X}^f} \right )\,.
\end{equation}
This action indicates how one can include the Newtonian potential in a manifestly reparametrization-invariant manner. It is relativistic (in the sense that the velocity does not have to be small), but not Lorentz invariant, as the potential breaks Lorentz symmetry. In the low-velocity expansion, the associated energy is $ m + \frac{1}{2} m v^2 + m \Phi  + m \Phi v^2 + ...$ where $\Phi v^2$ is the first post-Newtonian correction. Note, that for a particle moving on a geodesic, the potential and the velocity are not independent; thus, a weak Newtonian potential also implies a non-relativistic limit. 

To calculate the squeezing parameter, we need to calculate the second variation of the action in this geometry. We use Minkowski time to parameterize the particle worldline, $\tau \equiv t$. The presence of a background Minkowski spacetime, enables us to isolate tensor perturbations which are subject to \eqref{transversetraceless}; in fact the Newtonian potential itself can be regarded as a scalar perturbation. Then
the second variation \eqref{varact2} becomes 
\begin{equation}
     S_M^{(2)}
    =    \frac{m}{8} 
    \int  d t\,
\int \frac{d^3 k}{(2\pi)^3} \int \frac{d^3 k'}{(2\pi)^3}
\sum_{s s^\prime} 
\frac{
    q_{\vec k, s}(t) \epsilon_{i j}^{(s)} (\vec k)  \,
    q_{\vec k^\prime, s^\prime}(t) \epsilon_{l m}^{(s^\prime)} (\vec k^\prime) }{(-\bar{g}_{a b} \dot{X}^a \dot{X}^b)^{3/2}}
    e^{i (\vec{k} + \vec{k^\prime} )\cdot \vec{X} } 
    \,
    \dot{X}^i  \dot{X}^j
     \dot{X}^l   \dot{X}^m\,
    \label{s2varpoint}
\end{equation}
where we have exapanded the metric perturbations in a continuous set of Fourier modes.
The interaction Hamiltonian in the interaction picture is then
\begin{equation}
        H^I_{\text{int}} = - \frac{m}{4 M_p^2} \int \frac{d^3 k'}{(2\pi)^3} \int \frac{d^3 k'}{(2\pi)^3} \sum_{ss'} {\frac{1}{\sqrt{\omega_k \omega_{k'}}}}  \frac{\dot{X}^i  \dot{X}^j  \dot{X}^l  \dot{X}^m \epsilon^s_{ij}(\vec{k})  \epsilon^{s'}_{lm}(\vec{k}')}{\left( -\bar{g}_{ab} \dot{X}^a \dot{X}^b \right)^{\frac{3}{2}}}  \, e^{i(\vec{k} + \vec{k}') \cdot \vec{X}} \,
        \left(  a^{I \, \dagger}_{\vec{k},s}  a^{I \, \dagger}_{\vec{k'},s'} + a^I_{\vec{k},s} \, a^{I \, \dagger}_{\vec{k'},s'}  + h.c.\ \right)\,.
\end{equation}
The multi-mode squeezing parameter is in general a two-index object, and is dimensionful for continuous quantization. We can approximate it by a single mode squeezing operator by integrating over a relevant set of $\vec{k}'$ modes. 
Then, to leading order in the Magnus expansion, the (dimensionless) squeezing parameter is
\begin{equation}
\label{squeezing_matter}
        {\left( r \, e^{i \varphi} \right)}_{k,s,s'} = - \frac{i m}{2 \, M_p^2} \, \int \frac{d^3 k'}{(2\pi)^3} \,{\frac{1}{\sqrt{\omega_k \omega_{k'}}}} \int dt' \, \frac{\dot{X}^i \, \dot{X}^j \, \dot{X}^l \, \dot{X}^m}{\left( -\bar{g}_{ab} \, \dot{X}^a \, \dot{X}^b \right)^{\frac{3}{2}}} \, \epsilon^s_{ij}(\vec{k}) \, \epsilon^{s'}_{lm}(\vec{k}') \, e^{i(\vec{k} + \vec{k}') \cdot \vec{X}} \, e^{i (\omega_k + \omega_{k'}) t'}\,.
\end{equation}

Writing the wave vector as $k^i = k (\sin\theta \cos \phi, \sin \theta \sin \phi, \cos \theta)^i$, the two polarization tensors are
\begin{equation}
\epsilon_{ij}^{(+)} = \frac{1}{\sqrt{2}}
\begin{pmatrix}
\cos^2 \theta \cos^2 \phi - \sin^2 \phi & (1 + \cos^2 \theta) \sin \phi \cos \phi & -\frac{1}{2} \sin 2\theta \cos \phi \\
(1 + \cos^2 \theta) \sin \phi \cos \phi & \cos^2 \theta \sin^2 \phi - \cos^2 \phi & -\frac{1}{2} \sin 2\theta \sin \phi \\
-\frac{1}{2} \sin 2\theta \cos \phi & -\frac{1}{2} \sin 2\theta \sin \phi & \sin^2 \theta
\end{pmatrix},
\end{equation}
\begin{equation}
\epsilon_{ij}^{(\times)} = \frac{1}{\sqrt{2}}
\begin{pmatrix}
- \cos \theta \sin 2\phi & \cos \theta \cos 2\phi & \sin \theta \sin \phi \\
\cos \theta \cos 2\phi & \cos \theta \sin 2\phi & -\sin \theta \cos \phi \\
\sin \theta \sin \phi & -\sin \theta \cos \phi & 0
\end{pmatrix}. 
\end{equation}
Now, we can insert any geodesic trajectory, $X(t)$. Note that we cannot include non-geodesic trajectories without taking into account the agent responsible for the motion as that would violate the energy conservation, contradicting for example our choice of gauge.

At this point, one could insert any orbit here, including unbounded orbits; for unbounded trajectories the classical gravitational radiation emitted resembles Bremsstrahlung and was calculated in \cite{Kovacs:1977uw,Kovacs:1978eu}. 
Here, we simply consider a particle on a circular orbit in the equatorial plane
\begin{equation}
        X^a(t) \equiv \{t,R \cos (\omega \, t) , R \sin(\omega \, t) ,0  \}
\end{equation}
We suppose that at time $t=0$ the system is in its ground state. To leading order in the Newtonian potential
the squeezing parameter \eqref{squeezing_matter} is
\begin{equation}
        {\left( r \, e^{i \varphi} \right)}_{k,s,s'} = \int \frac{d^3 k'}{(2\pi)^3} c_{kk^\prime}\int_0^t dt'  \dot{X}^i
        \dot{X}^j \dot{X}^l \dot{X}^m
        \epsilon^s_{ij}(\vec{k})      \epsilon^{s'}_{lm}(\vec{k}')  \,e^{i R ( A  \cos{\omega t'} +  B  \sin{\omega t')}} e^{i (\omega_k + \omega_{k'}) t'}\,,
\end{equation}
where 
\begin{equation}
   c_{k k^\prime} = - \frac{i m}{8} \, \frac{4}{M_p^2 \, V} \,{\frac{{1-R^2 \omega^2 -3 \Phi \, \left( 1+R^2 \omega^2 \right)} }{\sqrt{\omega_k \omega_{k'}} \, \left( 1-R^2 \omega^2 \right)^{\frac{5}{2}}}}
   \approx
   - \frac{i m}{8} \frac{4}{M_p^2 \, V}  \,{\frac{{1 +\frac{3}{2} R^2 \omega^2 -3 \Phi  } }{\sqrt{\omega_k \omega_{k'}} }}
   \,, 
\end{equation}
\begin{equation}
        A = k \cos{\phi} \sin{\theta} + k' \cos{\phi'} \sin{\theta'}\,, \qquad
        B = k \sin{\phi} \sin{\theta}  + k'  \sin{\phi'} \sin{\theta'}\,,
\end{equation}
and 
\begin{align}
        &\dot{X}^i \epsilon^{+}(k)_{ij} \dot{X}^j = -\frac{R^2 \omega ^2 \left(\sin ^2\theta +\left(\cos ^2\theta +1\right) \cos (2 (\phi - \omega  t ))\right)}{2 \sqrt{2}} \,,  \\
        &\dot{X}^i \epsilon^{\times}(k)_{ij} \dot{X}^j = \frac{R^2 \omega ^2 \cos \theta  \sin (2 (\phi - \omega t  ))}{\sqrt{2}}\,.
\end{align}
To evaluate the integrals, it is useful to introduce
\begin{equation}
\label{ip}
I_p =    \int_0^t dt^\prime  e^{iR \, C \cos{(\omega t' - \alpha)} } \, e^{i (\omega_k + \omega_{k'}+p \omega) t'}\,,
\end{equation}
with $C = \sqrt{A^2 + B^2}$  and 
$\alpha = \arctan(\frac{B}{A})$ and where $p$ will be an integer.
With this, we find 
\begin{equation}
        {\left( r \, e^{i \varphi} \right)}_{k,\times \times} =  \int \frac{d^3 k'}{(2\pi)^3} \frac{c_{k k^\prime}R^4 \omega^4 \cos{\theta} \cos{\theta'}}{8} \big((e^{2i(\phi - \phi' )} + e^{-2i(\phi - \phi' )}) I_0 - e^{2i(\phi +\phi' )} I_{-4} - e^{-2i(\phi +\phi' )} I_4\big)
        \label{96}
\end{equation}
\begin{align}
    {\left( r \, e^{i \varphi} \right)}_{k,+ \times} &= -\int \frac{d^3 k'}{(2\pi)^3} c_{kk'} \frac{R^4 \omega^4}{16 i} \Big(2 \sin^2{\theta} \cos{\theta'} ( e^{2i\phi'} I_{-2} - e^{2i\phi'} I_2  ) \nonumber\\
    &~~~ + (\cos^2{\theta}+1) \cos{\theta'} \left(e^{2i(\phi+\phi^\prime)} I_{-4}
        - e^{-2i(\phi+\phi^\prime)} I_4 - (e^{2i(\phi-\phi')} -e^{-2i(\phi-\phi')}) I_0 \right)\Big)
\end{align}
\begin{align}
     {\left( r \, e^{i \varphi} \right)}_{k,++} &= \int \frac{d^3 k'}{(2\pi)^3} \frac{c_{kk^{\prime}} R^4 \omega^4}{32} \Big(4 \sin^2{\theta} \sin^2{\theta'} I_0  + 2\sin^2{\theta} \left( \cos^2{\theta'} + 1 \right) (e^{2i\phi' } I_{-2} + e^{-2i\phi' } I_2 ) \nonumber \\
     &+ 2 \left( \cos^2{\theta} + 1 \right) \sin^2{\theta'} (e^{2i\phi } I_{-2} + e^{-2i\phi } I_2) \nonumber \\
     &+ \left( \cos^2{\theta} + 1 \right) \left( \cos^2{\theta'} + 1 \right) ( (e^{2i(\phi-\phi')}- e^{-2i(\phi-\phi')}) I_0 + 
        e^{2i(\phi +\phi' )} I_{-4}+e^{-2i(\phi +\phi' )} I_4) \Big)
\end{align}
To proceed, let us examine the recurring integral \eqref{ip}. Using the identity
\begin{equation}
    e^{i z \cos{x}} = \sum^{\infty}_{n=-\infty} \, i^n J_n(z) e^{i n x}\,,
\end{equation}
we find that 
\begin{equation}
    I_p =
     \sum^{\infty}_{n=-\infty} \, i^n J_n(RC) e^{ -i n \alpha } \frac{e^{i (\omega_k + \omega_{k'}+p \omega + n \omega) t}-1}{i (\omega_k + \omega_{k'}+p \omega + n \omega)}
    \,.
\end{equation}
Notice that when $p+n$ is negative the denominator can vanish. For any given $\omega_k$, $\omega^\prime_k$, and $p$, there is (at most) one value of $n$ for which that can happen. Expanding the exponential for such cases we find that the fraction is just $t$. This linear growth of the squeezing parameter in time is familiar from the discussion of the harmonic oscillator undergoing parametric resonance at $\omega = 2 \omega_0$; here more generally the sum of $\omega_k$ and $\omega_{k^\prime}$ have to be an integer multiple of $\omega$, a phenomenon known as combination resonance. This also fixes the region of integration for $k'$.

The expressions above (prior to integration over $k'$) determine the squeezing parameter for all pairs of $k, k^\prime$ and for all pairs of polarization. Let us now identify the ones which are significant. As already mentioned, these are the ones for which there is linear growth in time, but in addition the angular dependence in the prefactors of the integrals also picks out certain angular combinations of $\vec{k}$ and $\vec{k}^\prime$. In particular, wherever there is a $\cos\theta$ or $\cos\theta^\prime$, the dominant contributions are at $0, \pi$ and whenever there is a $\sin\theta$ or $\sin\theta^\prime$, the dominant contributions are at $\pi/2$. 

Consider the $\times \times$ polarization; this is maximal at the poles $\theta, \theta^\prime = 0, \pi$ for which we can treat $\phi$ and $\phi^\prime$ as zero. Moreover, $C$ is also zero, so that the argument of all the Bessel functions is zero except for $J_0$, so that $n=0$ in the Bessel function expansions. But if $n=0$ and $p$ is positive, there is no way for the denominator in $I_p$ to vanish; thus the linear growth in time can only come from $I_{-4}$ so $\omega_k = 4 \omega- \omega_k^\prime$. Putting all this together we find
\begin{equation}
        |{ r}_{k,\times \times}| =  \frac{ m R^4 \omega^4}{16 \, M_p^2} \,(1 +\frac{3}{2} R^2 \omega^2 -3 \Phi  )
        \int \frac{d^3 k'}{(2\pi)^3}
       \frac{1} {\sqrt{(4 \omega- \omega_k^\prime) \omega_k^\prime} } \,
        t =  \frac{ 3 m R^4 \omega^6}{16 \pi \, M_p^2} \,(1 +\frac{3}{2} R^2 \omega^2 -3 \Phi  ) \, t 
        \label{961}\,,
\end{equation}
where in the integral $\omega_{k^\prime}$ ranges from 0 to $4 \omega$ and, for simplicity, we have just used $4 \pi$ for the angular integration.
Next, for the $\times +$ squeezing parameters, we find that the dominant contributions occur when $\theta, \theta^\prime = 0, \pi$ or $\theta^\prime = 0, \pi$ and $\theta = \pi/2$. In the first case we have
\begin{align}
    |{ r }_{k,k',+ \times}| &= \frac{ 3 \pi^2 m R^4 \omega^6}{2 \, M_p^2} \,(1 +\frac{3}{2} R^2 \omega^2 -3 \Phi  ) \, t 
\end{align}
This is the same as the $\times \times$ squeezing parameter. In the second case
\begin{align}
   |{ r }_{k,+ \times}| &= 
    \frac{m R^4 \omega^4}{16 M_p^2} \int \frac{d^3 k'}{(2\pi)^3} \,{\frac{{1 +\frac{3}{2} R^2 \omega^2 -3 \Phi  } }{\sqrt{\omega_k \omega_{k'}} }}
     (  I_{-2} -  I_2  )\,.
\end{align}
When $\theta = \pi/2$ and $\theta' = 0, \pi/2$, we have $C = k (= \omega_k)$ and $\alpha = \phi$ and hence the argument of the Bessel functions is $\omega_k R$. There are now various asymptotic forms, depending on whether $\omega_k$ and $\omega^\prime_k$ are large or small compared to $\omega$. If $\omega_k \ll \omega$, then the argument of the Bessel function is small (since we are in the non-relativistic limit $\omega R \ll 1$). Since $J_n(x) \approx \frac{(x/2)^n}{n!}$ for small $x$, the dominant terms arise for small $n$ i.e. for which $\omega^\prime_k$ is of the order of $\omega$. Alternatively, if $\omega_k \gg \omega$ (such that $\omega_k R \gg 1$), we have $J_n(x) \sim \sqrt{\frac{2}{\pi x}}$. That is, for $\omega_k R \ll 1$, we have 
\begin{align}
  |{ r }_{k,+ \times}| = 
    \frac{m R^4 \omega^4}{16 M_p^2}    \int \frac{d^3 k'}{(2\pi)^3} \,&{\frac{{1 +\frac{3}{2} R^2 \omega^2 -3 \Phi  } }{\sqrt{\omega_k \omega_{k'}} }} 
    \Big ( \frac{\omega_k R}{2}\delta_{\omega_k + \omega_k^\prime, \omega} + \delta_{\omega_k + \omega_k^\prime, 2 \omega} \nonumber \\
    &+
    \sum_{n=1}^\infty \big( \frac{\omega_k R}{2}  \big)^n 
    (\delta_{\omega_k + \omega_k^\prime, (n+2) \omega} - \delta_{\omega_k + \omega_k^\prime, (n-2) \omega}) \Big ) t
\end{align}
while, for $\omega_k R \gg 1$, we have
\begin{align}
  |{ r }_{k,+ \times}| &= 
    \frac{m R^4 \omega^4}{16 M_p^2} \, \int \frac{d^3 k'}{(2\pi)^3} \,{\frac{{1 +\frac{3}{2} R^2 \omega^2 -3 \Phi  } }{\sqrt{\omega_k \omega_{k'}} }} 
    \sqrt{\frac{\pi}{2 \omega_k R}} \sum_{n=1}^\infty \big ( \delta_{\omega_k + \omega_k^\prime, (n+2) \omega} -  \delta_{\omega_k + \omega_k^\prime, (n-2) \omega} \big ) t\,.
\end{align}
The $++$ squeezing parameters can be similarly calculated. There are four cases each with several terms, depending on whether $\theta, \theta^\prime$ are $0, \pi$ or $\pi/2$. The simplest case has $\theta, \theta^\prime$ both equal to 0 or $\pi$. Then
\begin{align}
  |{ r }_{k,+ +}| &= 
    \frac{m R^4 \omega^4}{16 M_p^2} \int \frac{d^3 k'}{(2\pi)^3} \,{\frac{{1 +\frac{3}{2} R^2 \omega^2 -3 \Phi  } }{\sqrt{\omega_k \omega_{k'}} }} \delta_{\omega_k + \omega_k^\prime, 4 \omega}\, t = \frac{3m R^4 \omega^6}{16 \pi M_p^2} \, \left( 1 +\frac{3}{2} R^2 \omega^2 -3 \Phi   \right) \, t \, .
\end{align}
Although the squeezing parameter grows with time, the amount of time available is not unbounded because the particle sheds energy via gravitational waves. One can estimate the maximum increase in $r$ as follows. A particle in a circular Newtonian orbit loses energy through quadrupole gravitational radiation,
\begin{equation}
\dot{E} = - \frac{32}{5} \frac{G^4 M^3 m^2}{c^5 R^5} \, ,
\end{equation}
where we have restored factors of $c$. Since the particle has total energy $E = -\frac{GMm}{2R}$, the amount of time spent at the orbit before the energy loss becomes comparable to the particle's energy is roughly
\begin{equation}
\Delta t \sim \frac{E}{\dot{E}} \sim \frac{c^5 R^4}{G^3 M^2 m}
\end{equation}
and hence
\begin{equation}
\Delta r \sim \frac{m G R^4 \omega^6}{c^7} \Delta t \sim \frac{c^5 R^4}{G^3 M^2 m} \sim \frac{R_s}{R} \,,
\end{equation}
where $R_s = \frac{2GM}{c^2}$ is the Schwarzschild radius. The innermost circular orbit (for which $\frac{R_s}{R} = 1/3$) therefore provides an upper bound for the squeezing parameter. At that location, however, our calculations are subject to nontrivial post-Newtonian corrections.

The formalism presented here, which is both general and powerful, can be readily applied to other gravitating systems. It would be very interesting to find non-cosmological scenarios for which large squeezing parameters are generated by realistic astrophysical sources.

\section*{Acknowledgements}
\noindent
We would like to thank Lars Aalsma, Giordano Cinta,
and George Zahariade for helpful conversations.
RW and MP acknowledge support from the Heising-Simons Foundation under the ``Observational Signatures of Quantum Gravity'' collaboration grant 2021-2818 and the U.S. Department of Energy, Office of High Energy Physics,
under award DE-SC0019470.
RW also acknowledges support from the STFC consolidated grant ST/X000583/1 ``New Frontiers in Particle Physics, Cosmology and Gravity'' in the final stages of this project.
FW is supported by the U.S. Department of Energy under award DE-SC0012567 and by the Swedish Research Council under Contract No. 335-2014-7424.

\appendix

\section{Coherent and Squeezed States in Quantum Mechanics}
\label{app:coh}
In this appendix we review properties of coherent and squeezed states in quantum mechanics for the convenience of the reader.
\subsection{Coherent States}
Coherent states saturate the Heisenberg minimum uncertainty bound and expectation values of operators follow classical trajectories.  They are sometimes regarded to be the most classical quantum states \cite{Loudon:1987pxc}. Here, we recall some of their basic properties.

Coherent states $\ket{\alpha}$ are eigenstates of the annihilation operator with eigenvalue $\alpha$ 
\begin{equation}
    a \, \ket{\alpha} = \alpha \, \ket{\alpha}
\end{equation}
and can be obtained by acting with the displacement operator 
\begin{equation}
     D(\alpha) = \exp{\big( \alpha \, a^{\dagger} - \alpha^* \, a \big)}\,,
\end{equation}
where $\alpha \in \mathbb{C} $, on the vacuum of $a$, i.e.\ $\ket{\alpha} = D(\alpha) \, \ket{0}$.
From a Lagrangian perspective, such states can be generated from harmonic oscillators by adding a source term $J(t) x$, i.e.\ 
\begin{equation}
\label{Lagharmapp}
        L = \frac{1}{2} {\dot x}^2 - \frac{1}{2} \omega_0^2 x^2 + J(t) x \,.
\end{equation}
The associated interaction Hamiltonian to this Lagrangian is $H_{\text{int}} = -J(t) \, x$. 
Quantizing as \eqref{quantize} and using Baker-Campbell-Hausdorff one easily shows that the state 
\begin{equation}
    \ket{\psi } = T \, \big(e^{- i \int H^I_{\mathrm{int}}(t') dt'} \, \big)\, \ket{0} 
     = T \, \big( \, e^{\frac{i}{\sqrt{2 \omega_0}} \int_0^t J(t')\, (a \, e^{-i \omega_0 t'} + a^{\dagger} \,e^{i \omega_0 t'})dt'} \, \big) \, \ket 0
\end{equation}
is a coherent state
\begin{align}
       a^{I}(t) \ket \psi  = a e^{-i \omega_0 t} \ket \psi = \frac{i}{\sqrt{2 \omega_0}} \int_0^t J(t_1) dt_1 e^{i \omega_0 (t_1-t)} \ket \psi = \alpha \ket \psi \,.
\end{align}

The fact that the expectation values follow the classical trajectories can be seen as follows. Applying Weinberg's formula \cite{Weinberg:2005vy}, which states that for an operator $Q$ the expectation value reads
\begin{equation}
\label{weinberg}
        \braket{Q(t)} 
    = \sum_{n = 0}^\infty i^N
    \int_{-\infty}^t dt_N 
    \int_{-\infty}^{t_N} dt_{N-1}
    \int_{-\infty}^{t_2} dt_{1}
    \langle \, 0 | \, {[H^I_{\text{int}}(t_1), [H^I_{\text{int}}(t_2), ...
    [H^I_{\text{int}}(t_N),Q^I(t)]
    ]]} \, | \, 0 \, \rangle\,,
\end{equation}
one easily computes $\braket{x(t)}$:
For $H^I_{\text{int}}(t)$ given above, the commutator $[\, H^I_{\text{int}}(t_1),x^I\,]$
is a c-number and thus
\begin{equation}
        \braket{x(t)}
        = \int_{-\infty}^t dt_1 \, J(t_1) 
        \frac{\sin(\omega_0 (t-t_1))}{\omega_0}\,,
\end{equation}
which is precisely the classical solution to the equations of motion $\ddot{x} + \omega_0^2 x = J$ of \eqref{Lagharmapp}. 

\subsection{Squeezed states}
\begin{figure}[h!]
    \centering
    \subfigure{
        \includegraphics[width=0.489\textwidth]{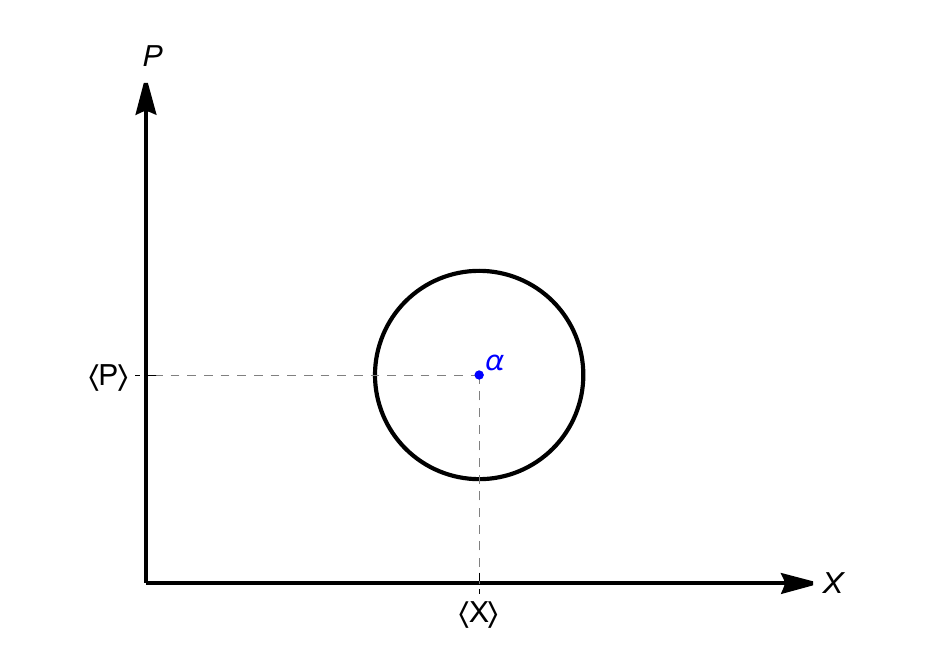}}
    \hfill
    \subfigure{
\includegraphics[width=0.489\textwidth]{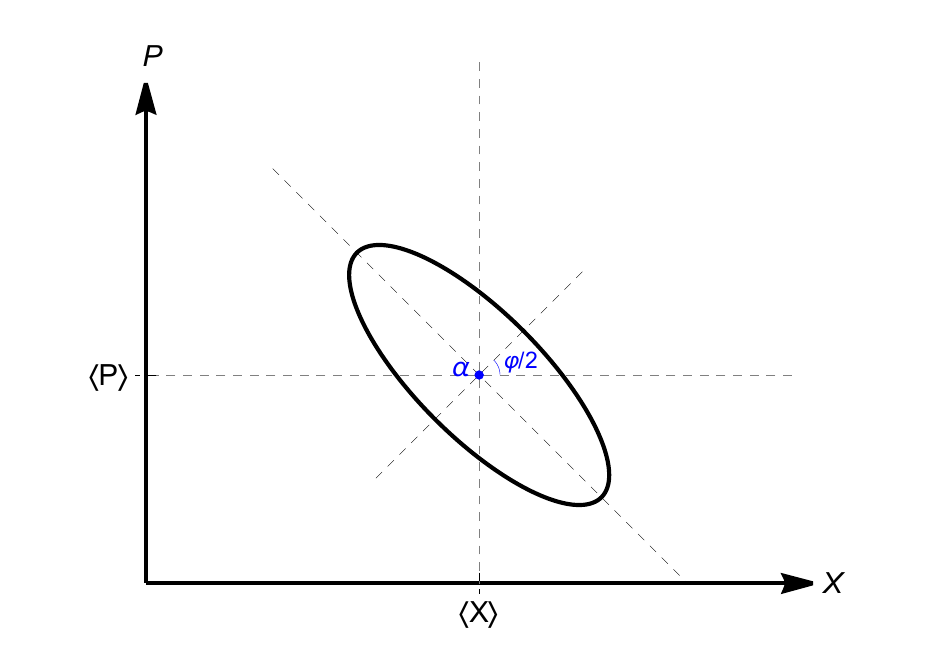}}
    \caption{Phase space diagram for a coherent state $D(\alpha)\ket{0}$ (left) and a displaced squeezed state $D(\alpha) S(r, \varphi) \ket{0}$ (right). The angle $\varphi/2$ determines the direction of reduced uncertainty in phase space. The real and imaginary parts of $\alpha$ correspond to the expectation values of $X$ and $P$.
    }
    \label{fig6}
\end{figure}

Squeezed states of the harmonic oscillator are obtained by acting on the ground state of the annihilation operator $ \ket{0}$ with an exponential operator containing terms quadratic in the creation and annihilation operators. Indeed, such states can be obtained by acting with $S(r, \varphi)$ on $ \ket{0}$ where
\begin{equation}
    S(r,\varphi) = \exp{\left( \frac{r}{2} \, \left( e^{-i \, \varphi} \, a^2 - e^{i \, \varphi} \, {(a^{\dagger})}^2 \right) \right)} 
\end{equation}
and $0 \le r < \infty$ is the squeezing parameter and $0 \le \varphi \le 2 \pi$ is the squeezing angle.
Squeezed states share an important property with coherent states: when the squeezing angle is set to zero, they  saturate the Heisenberg minimum-uncertainty bound.
Their key difference lies in how the uncertainty is distributed among  the phase space quadratures $X = \frac{1}{2} \, \left( a+a^{\dagger} \right) = x \sqrt{\frac{ m \omega_0}{2}}$, $P=\frac{1}{2i} \, (a-a^{\dagger}) = 
\frac{p}{\sqrt{2 m \omega_0}}
$. In a coherent state the variances of the quadratures are equal, whereas in a squeezed state one quadrature has reduced uncertainty while the conjugate quadrature has a correspondingly increased uncertainty, i.e.\
\begin{equation}
    \left( \Delta X \right)^2 = \frac{1}{4} \, \left( e^{-2r} \, \cos^2{\frac{\varphi}{2}} + e^{2r} \, \sin^2{\frac{\varphi}{2}} \right) \, , \quad
    \left( \Delta P \right)^2 = \frac{1}{4} \, \left( e^{-2r} \, \sin^2{\frac{\varphi}{2}} + e^{2r} \, \cos^2{\frac{\varphi}{2}} \right)\,.
\end{equation}
By increasing the squeezing parameter, the variance of one quadrature can be made arbitrarily small at the cost of enlarging the variance of the other. Figure \ref{fig6} shows the phase space uncertainty of a displaced squeezed state  (red ellipse)  and the one of a coherent state (black circle).

\section{Coherent states in gravity}
The body of the paper shows that time‑dependent quadratic couplings generically produce squeezed graviton states. Here we show that linear coupling to sources, leads to coherent states.
We consider gravitational perturbations on Minkowski in the presence of matter and we assume that the second variation of the matter Lagrangian vanishes. The first and second variation of the action then reduce to
\begin{equation}
  S^{(1)}[ \bar{g},h] +      S^{(2)}[ \bar{g},h] = \frac{1}{8 \kappa} \, \int \, d^4x \, h^{ab} \, \left(\bar{\Box} \, h_{ab} - \frac{1}{2} \, \eta_{ab} \, \bar{\Box} h^{c}_c + 4 \kappa T_{ab}
   \right)\,.
   \label{action1plus2}
\end{equation}
and the associated equations of motion for the perturbation read
\begin{equation}
\bar{\Box} \, h_{ab} - \frac{1}{2} \, \eta_{ab} \, \bar{\Box} h^{c}_c = -2 \kappa \, T_{ab}\,.
\end{equation}
These are solved by 
\begin{equation}
\label{classh}
        h^{class}_{a b}(t,\vec{x}) = \frac{\kappa}{2\pi}
    \int
     d^3 y \, \, 
    \frac{T_{a b} (t- |\vec{x} - \vec{y}|, \vec{y}) - \frac{1}{2} \, \eta_{a b} \, T^c_c (t- |\vec{x} - \vec{y}|, \vec{y})}{|\vec{x} - \vec{y}|}\,.
\end{equation}

 We now work in interaction picture, where we 
 promote $h_{a b}^I$ to an operator and  consider the state 
\begin{equation}
    \ket{\alpha} = T \, \exp{\left( - i \int_{-\infty}^t \, H^I_{\text{int}}(t) \, dt \right)} \, \ket{0}\,,
\end{equation}
where 
\begin{equation}
   H^I_{\text{int}}(t) = -{\frac{1}{2}}\int T^{a b}(t, \vec x) h^I_{a b}(t, \vec x) d^3 x
   \label{HIeq}
 \end{equation}
 is the interaction Hamiltonian.
By expanding the perturbation in creation and annihilation operators, one can easily check that this state is an eigenstate of the annihilation operator by generalizing the calculations in the appendix \ref{app:coh}. Here, we simply check that its expectation value follows the classical trajectory, i.e.\
$\bra{\alpha}{{h}}_{a b} \ket{\alpha} = h^{class}_{a b}$.
Using \eqref{weinberg}, the nested commutator truncates at first order since the commutator of two $h_{ab}$ is a $c$-number
\begin{align}
     \bra{\alpha}{{h}}_{a b}(t, \vec x) \ket{\alpha} &= -{\frac{i}{2}} 
      \int_{-\infty}^t  d\tilde t\int d^3  y\, T^{c d}(\tilde t, \vec y)\, \, \bra{0}
      [{h}^I_{c d}(\tilde t, \vec y), {h}^I_{a b}(t, \vec x)] \ket{0} \nonumber\\
      &=   {\frac{i}{2}}
      \int_{-\infty}^{\infty}  d\tilde t\int d^3 y\, T^{c d}(\tilde t, \vec y) D^{R}_{c d a b}(\vec x-\vec y) \,,
      \label{expechab}
\end{align}
where we identify the retarded graviton propagator as
\begin{align}
D^{R}_{c d a b} (\vec x- \vec y) &= \Theta(t-\tilde{t}) \bra 0 [{h}^I_{a b}(t, \vec x),{h}^I_{c d}(\tilde{t}, \vec y)] \ket 0\,. 
\end{align}
 The momentum space retarded propagator for a massless graviton in the harmonic gauge $\partial^\mu h_{\mu \nu} - \frac{1}{2} \partial_\nu h^c_c = 0$ reads \cite[cf.\ (2.54)]{Hinterbichler:2011tt} 
\begin{equation}
\label{hinter}
    \tilde{D}_{a b c d} = -\frac{2 i \kappa }{p^2}  \left( \, \eta_{a d} \eta_{b c} + \eta_{a c} \eta_{b d} - \eta_{c d} \eta_{a b} \, \right)\,.
\end{equation}
Fourier transforming \eqref{hinter} and using that
\begin{equation}
    \int_{\mathbb{R}^4} \frac{d^4 p}{{(2 \pi)}^4 p^2} e^{i p_0  (t- \tilde t) - i\vec{p} \cdot  (\vec{x}-\vec{y})}
    = \frac{1}{4 \pi |\vec x- \vec y|} \left( \, \delta(t-\tilde{t} + |\vec x- \vec y|) - \, \delta(t-\tilde{t} - |\vec x- \vec y|)\right)\,.
\end{equation}
one readily obtains
\begin{equation}
    \langle {h}_{a b} (t,\vec{x}) \rangle  = \frac{\kappa }{4 \pi} \, \int d^3 y \frac{T^{c d}(t-|\vec x- \vec y|, \vec y)}{|\vec x-\vec y|} \,  \left( \, \eta_{a d} \eta_{b c} + \eta_{a c} \eta_{b d} - \eta_{c d} \eta_{a b} \, \right)
\end{equation}
and thus 
\begin{equation}
    \langle {h}_{a b} (t,\vec{x}) \rangle =  \frac{\kappa}{2 \pi} \int d^3 y \, \, \frac{T_{a b}(t- |\vec{x} - \vec{y}|, \vec{y}) - \frac{1}{2} \, \eta_{a b} \, T^c_c (t- |\vec{x} - \vec{y}|, \vec{y})}{|\vec{x} - \vec{y}|}\,,
\end{equation}
which is exactly equal to the classical value \eqref{classh}. 
\bibliographystyle{JHEP-2}
\bibliography{references}
\end{document}